\documentclass[acmsmall]{acmart}

\usepackage[ruled,vlined,linesnumbered]{algorithm2e}
\usepackage{amsmath}
\usepackage{multirow}
\usepackage{wrapfig}
\usepackage{rotating}

\setcopyright{rightsretained}
\acmJournal{PACMHCI}
\acmYear{2024} 
\acmVolume{8}
\acmNumber{ETRA}
\acmArticle{228}
\acmMonth{5}
\acmDOI{10.1145/3655602}

\received{November 2023}
\received[revised]{January 2024}
\received[accepted]{March 2024}

\graphicspath{ {./low_res/} }

\usepackage{array}
\newcolumntype{C}[1]{>{\centering\arraybackslash}m{#1}}

\makeatletter
\gdef\@copyrightpermission{
\begin{minipage}{0.2\columnwidth}
 \href{https://creativecommons.org/licenses/by/4.0/}{\includegraphics[width=0.90\textwidth]{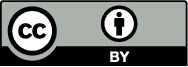}}
\end{minipage}\hfill
\begin{minipage}{0.8\columnwidth}
 \href{https://creativecommons.org/licenses/by/4.0/}{This work is licensed under a Creative Commons Attribution International 4.0 License.}
\end{minipage}
\vspace{5pt}
}
\makeatother

\begin{document}

\title{Impact of Design Decisions in Scanpath Modeling}

\author{Parvin Emami}\email{parvin@emami@uni.lu}
\affiliation{%
  \institution{University of Luxembourg}
  \country{Luxembourg}
}

\author{Yue Jiang}\email{yue.jiang@aalto.fi}
\author{Zixin Guo}\email{zixin.guo@aalto.fi}
\affiliation{%
  \institution{Aalto University}
  \country{Finland}
}

\author{Luis A. Leiva}\email{name.surname@uni.lu} 
\affiliation{%
  \institution{University of Luxembourg}
  \country{Luxembourg}
}

\renewcommand{\shortauthors}{Parvin Emami, Yue Jiang, Zixin Guo, \& Luis A. Leiva}

\begin{abstract} 
Modeling visual saliency in graphical user interfaces (GUIs) allows to understand how people perceive GUI designs and what elements attract their attention. One aspect that is often overlooked is the fact that computational models depend on a series of design parameters that are not straightforward to decide. We systematically analyze how different design parameters affect scanpath evaluation metrics using a state-of-the-art computational model (DeepGaze++). We particularly focus on three design parameters: input image size, inhibition-of-return decay, and masking radius. We show that even small variations of these design parameters have a noticeable impact on standard evaluation metrics such as DTW or Eyenalysis. These effects also occur in other scanpath models, such as UMSS and ScanGAN, and in other datasets such as MASSVIS. Taken together, our results put forward the impact of design decisions for predicting users' viewing behavior on GUIs.
\end{abstract}

\ccsdesc[500]{Human-centered computing~Empirical studies in ubiquitous and mobile computing}
\ccsdesc[300]{Computing methodologies~Computer vision}

\newcommand\DELETE[1]{\textcolor{red}{#1}}
\newcommand\ADD[1]{\textcolor{black}{#1}}
\newenvironment{delete}{\par\color{red}}{\par}
\newenvironment{add}{\par\color{black}}{\par}

\newcommand{\loss}{\mathcal{L}}
\newcommand{\image}{\mathcal{I}}
\newcommand{\encoder}{\mathcal{E}}
\newcommand{\decoder}{\mathcal{D}}
\newcommand{\normal}{\mathcal{N}}

\keywords{Visual Saliency; Interaction Design; Computer Vision; Deep Learning; Eye Tracking}
 
\begin{teaserfigure}
  \Description{}
  \label{fig:teaser}
\end{teaserfigure}

\maketitle

\section{Introduction}

Understanding how user attention is allocated in graphical user interfaces (GUIs) is an important research challenge, 
considering that many different GUI elements (e.g. buttons, headers, cards, etc.) may stand out and engage users effectively~\cite{terzimehic2019review}. 
By modeling eye movement patterns of visual saliency, 
we can gain invaluable insights into how users perceive and interact with GUIs,
without having to recruit users early in the GUI design process.
When presented with a GUI screenshot, a saliency model can predict how users would spend their attention, 
typically over short periods of time during free-viewing scenarios (bottom-up saliency)
or over longer periods in task-based scenarios (top-down saliency).
We can use these predictions to quantify the impact of a visual change in the GUI  
(e.g. after rescaling an element or changing its position),
optimize some design components so that it can grab less or more user attention,
or understand whether users notice some element after some time of exposure. 

Saliency models can predict either (static) saliency maps~\cite{zhu2019prediction} or (dynamic) scanpaths~\cite{kummerer2021state}.
Most research has focused on predicting 
saliency maps~\cite{jiang2023ueyes, fosco2020predicting, cornia2018sam, kummerer2022deepgaze, assens2018pathgan}, 
overlooking key temporal aspects like fixation timing and duration. 
Saliency maps show aggregated fixation locations, i.e., areas that users will pay attention to, where the eye remains relatively static. 
In contrast, scanpaths contain sequential information on individual fixations and sometimes also saccades, 
i.e., eye movements between those points, thus retaining information about fixation order and their temporal dynamics. 
In other words, scanpaths comprise rich data from which second-order representations such as saliency maps can be computed. 
Critically, scanpaths can inform about the users' visual flows, 
through which one can better assess how attention deploys over time. 
This is vital for understanding how individual users, instead of a group of users, would perceive the GUI 
and for design adjustments that encourage viewing GUI elements in a desired order for different people~\cite{eraslan2016eye}.
For these reasons, in this paper we focus on scanpath models of visual saliency.

One problem that is often overlooked is that many computational models of visual saliency rely on a set of \emph{design parameters} 
that must be defined beforehand.
Some of them can be inferred from the collected data, such as deciding the number of fixations to predict on average.
However other parameters must be established by the researcher, 
such as deciding the resolution of the GUI screenshots for model input.
These design parameters cannot be learned e.g. through backpropagation, 
so researchers have to rely on their own expertise, trial and error, previous work, or best practices.
To the best of our knowledge, their impact on downstream performance has not been systematically analyzed.
We believe this kind of analysis is very much needed because any evaluation depends on the quality of the model predictions,
so it may be the case that small variations on some parameters produce different performance results.
In this context, we pose the following research question:
\textbf{To what extent do saliency model predictions depend on the choices made in their design parameters?}

We use DeepGaze++~\cite{jiang2023ueyes} as a reference model 
to investigate the potential impact that different design parameters may have in scanpath prediction.
DeepGaze++ is a state-of-the-art scanpath model for visual saliency prediction that has shown promising results. 
However, like other models, it relies on ``hardcoded'' design parameters such as the aforementioned input screenshot size or, more interestingly, the masking radius used for inhibition of return (IOR) mechanisms. IOR, which refers to the phenomenon where attention is less likely to return to previously attended locations, plays a crucial role in visual perception. If the masking radius is not appropriately calibrated, there is a possibility of omitting potentially salient areas within a GUI~\cite{klein2000inhibition}.
Also, as explained later, DeepGaze++ relies on a sub-optimal IOR weighting mechanism limited to 12 fixation points, 
so we propose a new weighting scheme to overcome this limitation as an aside research contribution.

While using hardcoded design parameters may be the most straightforward approach, 
determining their optimal values for each type of GUI remains an open question. 
For example, if we were to have a masking radius equivalent to the whole image size, 
the model could only predict one fixation, as the whole GUI would have been masked out
(i.e. no other GUI parts could be fixated on because, by definition, 
there is nothing left to fixate on if everything is masked out).
In this paper, we focus on three \emph{key} design parameters common to every scanpath model:
\begin{description}
    \item[Input image size,] which determines the granularity of the predicted fixations 
    (higher resolution gives more room to fixate on more GUI parts).
    \item[IOR decay,] which implements the importance that previous fixations have in successive fixation predictions. 
    \item[Masking radius,] which allows to prevent that previously fixated GUI parts are fixated on again.
\end{description}
We study the impact of these parameters on different GUI types, 
including web-based, mobile, and desktop applications. 
Then, we show that the optimized parameters we discovered for DeepGaze++~\cite{jiang2023ueyes} 
help improve the performance of other scanpath models
and also generalize to other datasets.
In sum, this paper makes the following contributions:
\begin{enumerate}
    \item A comparative study on model design parameters on scanpath prediction performance, 
    across different types of GUIs.
    \item An optimized set of design parameters for scanpath models, evaluated across multiple models and datasets.
    \item A new IOR decay parameter, designed to work with an arbitrary number of fixation points.
\end{enumerate}
\section{Related work}

Visual saliency prediction in GUIs has witnessed substantial growth in recent years~\cite{yan2021review, zhu2021viewing, yang2019predicting}, 
driven by the increasing demand for accurate models that can anticipate where human attention is likely to be directed within digital displays. 
As previously hinted, existing research has predominantly focused on the development and refinement of visual saliency prediction models, 
often overlooking the crucial impact that different design parameters may have.
For example, \citet{de2022scanpathnet} proposed ScanpathNet, a deep learning model inspired by neuroscience,
and noted that model performance was influenced by the choice of the number of Gaussian components.
This highlights the significant impact that the choice of design parameters may have. 

Previous research considered different image sizes to predict visual saliency, 
informed by the datasets they used for training the models~\cite{kummerer2021state, bao2020human, chetouani2020use}.
It became apparent that higher image resolutions are not desired, as drifting errors tend to increase~\cite{li2018learning},
but no systematic examination was provided in this regard.
In addition, \citet{parmar2022aliased} demonstrated that generative models 
are particularly sensible to image resizing artifacts such as quantization and compression.
Therefore, we decided to examine the impact of image resizing in visual saliency prediction. 

Generating fixation sequences accurately, while promoting a coherent and natural order, 
remains as the main challenge in scanpath modeling. 
\citet{itti1998model} introduced the Inhibition of Return (IOR) mechanism as a way to ensure that predicted fixations do not bounce back and forth around previously visited areas. This was later exploited in scanpath modeling~\cite{bao2020human, sun2019visual}, although all scanpath models that incorporate IOR employ a fixed masking radius~\cite{chen2018scanpath, david2019predicting}. Furthermore, there is no discussion about how this radius affects model predictions. As mentioned in the previous section, when this radius is too large, the model's ability to predict multiple fixation points will be severely limited. Therefore, we decided to examine the impact of masking radius and IOR decay in visual saliency prediction. 

Several methods, including deep neural networks and first-principle models, 
have been proposed to predict scanpaths in natural images~\cite{cornia2018sam}, videos~\cite{li2023scanpath}, and, more recently, GUIs~\cite{jiang2023ueyes}. 
\citet{ngo2017saccade} developed a recurrent neural network to predict sequences of saccadic eye movements
and \citet{wloka2018active} predicted fixation sequences by relying on a ``history map'' of previously observed fixations.
These works were evaluated on small datasets and using a limited set of metrics,
therefore it remains unclear whether these models can compare favorably in GUIs. 

Later on, \citet{xia2019predicting} introduced an iterative representation learning framework to predict saccadic movements.
More recently, \citet{jiang2023ueyes} developed DeepGaze++ based on DeepGaze~III~\cite{kummerer2022deepgaze}.
DeepGaze~III takes both an input image and the positions of the previous fixation points to predict a probabilistic density map for the next fixation point. It frequently tends to predict clusters of nearby points, potentially leading to stagnation within those clusters. To address this problem, DeepGaze++ recurrently chooses the position with the highest probability from the density map, concurrently implementing a custom IOR decay to suppress the selected position in the saliency map. (As explained in the next section, this decay only works for a relatively small number of fixation points, therefore we propose a new IOR decay to address this limitation.) Nevertheless, DeepGaze++ is a state-of-the-art scanpath model so we use it in our investigation.
\section{Methodology}

The goal of scanpath prediction is to generate a plausible sequence of fixations,
where fixations refer to distinct focal points during visual exploration. 
As previously mentioned, our study leverages the advanced capabilities of DeepGaze++ to answer our research question.

\subsection{Dataset}

In our study, we use the UEyes dataset~\cite{jiang2023ueyes}, 
a collection of eye-tracking data over 1,980 screenshots covering four GUI types (495 screenshots per type): 
posters, desktop UIs, mobile UIs, and webpages. 
This dataset was collected from 66 participants (23\,male, 43\,female) aged 27.25 years (SD=7.26)
via a high-fidelity in-lab eye tracker Gazepoint GP3. 
Participants had normal vision (43) or wore either glasses (18) or contact lenses (5).
No participant was colorblind.
Eye-tracking data were recorded after participant-specific calibration, 
a step that accounts for variables
such as eye-display distance and visual angle,
to ensure accurate recording of eye data. 
Participants were given 7 seconds to freely view each GUI screenshot in a $1920$x$1200$\,px monitor.
For our study, we considered the same data partitions as in the UEyes dataset: 
1,872 screenshots for training and 108 for testing.
Our experiments are performed over the training partition of the UEyes dataset.
The testing partition simulates unseen data, therefore it is used for final model evaluation.

\subsection{Design parameters}

In the following, we describe the parameters we have considered for our study.
Note that they are all common to every scanpath model, they cannot be inferred from data, and they cannot be learned automatically.
In our experiments, whenever we modify the values of each parameter, everything else remains constant.
This way, it is easy to understand the concrete influence of each parameter in model performance.

\subsubsection{Image size}

Each GUI type has a preferred size (e.g. desktop applications are usually designed for FullHD monitors)
or proportion (e.g. mobile apps have around 9:16 aspect ratio).
When GUI images are resized (downsampled), to speed up computations, the models may perform differently.
Therefore, it is unclear which image resolution should be used as model input.
To shed light in this regard, we tested different input sizes and aspect ratios. 

\subsubsection{IOR decay}

Initially introduced by \citet{posner1984components}, 
IOR is a neural mechanism that suppresses visual processing within recently attended locations. 
In the context of scanpath modeling, 
DeepGaze++ uses an IOR decay of $\displaystyle 1 - 0.1 (n - i - 1)$, 
for the $i${-th} fixation point when predicting $n$ fixation points, 
to prevent that older fixation points are likely to be revisited.
As can be noticed, this is limited to a maximum number of 12 fixation points,
after which the decay values may become negative;
e.g., for $n=13$ the first fixation point $i=1$ gets an IOR of $-0.1$. 
Consequently, we have developed a new IOR decay designed to accommodate any number of fixation points. 
We propose $\displaystyle \gamma^{(n - i - 1)}$, 
in which $\gamma$ is a design parameter, between 0 and 1, that we also analyze systematically.

\subsubsection{Masking radius}

To implement any IOR mechanism, we need to mask some areas around the previous fixation points. 
However, determining the optimal size of the masked areas is unclear.
Therefore, we consider the masking radius as a third design parameter and, 
in line with the previous discussions, 
examine how various masking radii impact the scanpath prediction results.

\subsection{Evaluation metrics}

We employ a set of four metrics that, together, 
provide a holistic assessment about the predictive performance of scanpath models~\cite{mathot2012simple, fahimi2018sequential, anderson2015comparison}:
Dynamic Time Warping (DTW), 
Eyenalysis,
Determinism,
and Laminarity.
These metrics are well-established in the research literature~\cite{fahimi2021metrics}. 
While DTW measures the location and sequence of fixations in temporal order,
Eyenalysis measures only locations and Determinism measures only the order of fixation points. 
Conversely, Laminarity is a measure of repeated fixations on a particular region,
without considering their location or order.
\autoref{tbl:metrics} provides an overview of these metrics.

\begin{table}[!ht]
    \centering
    \begin{tabular}{lll}
    \toprule
        \textbf{Metric} & \textbf{Location} & \textbf{Order} \\
    \midrule
        DTW         & Yes & Yes \\
        Determinism & No  & Yes \\
        Eyeanalysis & Yes & No \\
        Laminarity  & No  & No \\
    \bottomrule
    \end{tabular}
    \caption{Overview of the chosen scanpath evaluation metrics~\protect\cite{fahimi2021metrics}.}
    \label{tbl:metrics}
\end{table}

\subsubsection{Dynamic Time Warping (DTW)} First introduced by \citet{berndt1994using}, DTW is a method for comparing time series with varying lengths. It involves creating a distance matrix between two sequences and finding the optimal path that respects boundary, continuity, and monotonicity conditions. 
The optimal solution is the minimum path from the starting point to the endpoint of the matrix. 
DTW identifies such an optimal match between two scanpaths in an iteratively manner, ensuring the inclusion of critical features~\cite{muller2007dynamic, wang2023improved}.


\subsubsection{Eyenalysis} This is a technique that involves double mapping of fixations between two scanpaths, aiming to reduce positional variability~\cite{mathot2012simple}. Like in DTW, this approach may result in multiple points from one scanpath being assigned to a single point in the other. Eyeanalysis performs dual mapping by finding spatially closest fixation points between two scanpaths, measuring average distances for these corresponding pairs.


\subsubsection{Determinism} This metric gauges diagonal alignments among cross-recurrent points, representing shared fixation trajectories~\cite{fahimi2021metrics}. With a minimum line length of $L = 2$ for diagonal elements, Determinism measures the congruence of fixation sequences. Computed as the percentage of recurrent fixation points in sub-scanpaths, Determinism considers pairs of distinct fixation points from two scanpaths, enhancing the original (unweighted) Determinism metric for subscanpath evaluation.

\subsubsection{Laminarity} 
It measures the percentage of fixation points on sub-scanpaths in which all the pairs of corresponding fixation points are recurrences but all such recurrent fixation point pairs contain the same fixation point from one of the scanpaths~\cite{anderson2015comparison, fahimi2021metrics}. 
In sum, Laminarity indicates the tendency of scanpath fixations to cluster on one or a few specific locations.


\section{Experiments}

In the following, we report the experiments aimed at finding the optimal set of design parameters.
For the sake of conciseness, we consider DTW for determining the best result for each design parameter,
as this metric accounts for both location and order of fixations (\autoref{tbl:metrics}).


\subsection{Sensitivity to input image size} 

\begin{figure}[!t]
  \centering
  \def\w{0.4\linewidth}
  \begin{tabular}{ cccc } 
    \includegraphics[width=0.25\linewidth]{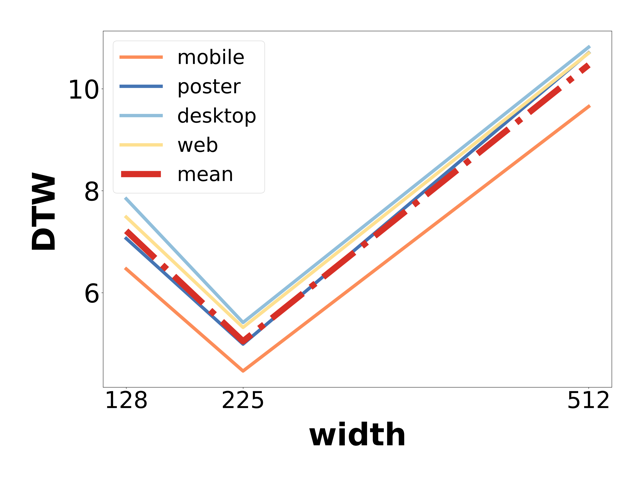}
    \includegraphics[width=0.25\linewidth]{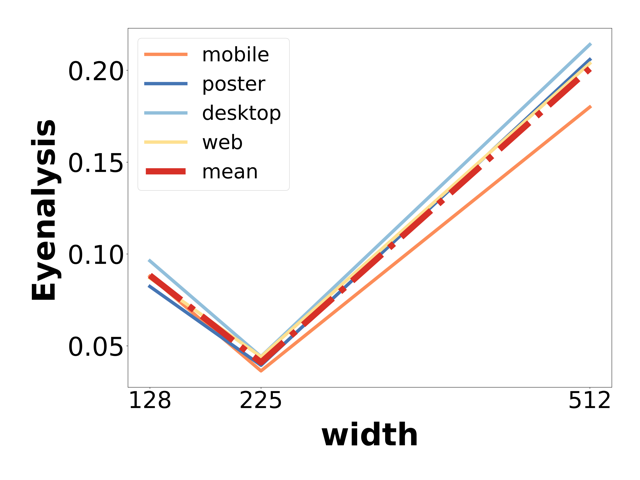} 
    \includegraphics[width=0.25\linewidth]{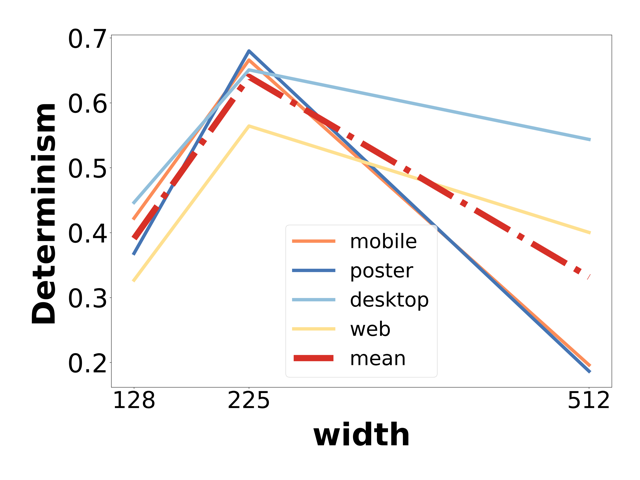} 
    \includegraphics[width=0.25\linewidth]{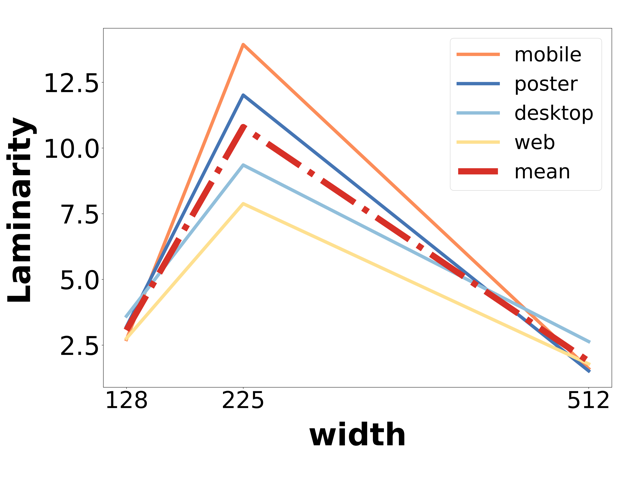}
    \\ 
  \end{tabular}
  \caption{
    Impact of resizing to square or non-square image on different GUI types. 
      The height is always fixed to 225\,px. 
      We consider widths of 128, 225, and 512\,px. 
      The best results are observed for widths of 225\,px (resulting in a square aspect ratio).
    }
  \label{fig:sen-square-vs-nonsquare}
\end{figure}

We analyzed the impact of resizing under different aspect ratios (square and non-square images).
In the first experiment (\autoref{fig:sen-square-vs-nonsquare}), 
the height of the resized images remained constant at 225\,px, 
as suggested in previous studies~\cite{jiang2023ueyes}, 
while we modified their width.
The other width values were chosen as the closest powers of two
around this baseline value of 225, for convenience.

The results, presented in \autoref{fig:sen-square-vs-nonsquare}, 
indicate that resizing any input image to a square aspect ratio
consistently yields superior performance across all GUI types. 
An intriguing observation is that mobile GUIs 
are particularly sensible to this parameter
as compared to other GUI types.
We attribute this effect to the fact that mobile GUIs,
despite having the largest aspect ratio,
make heavy use of icons and usually icons have a square aspect ratio.

\begin{figure}[!t]
  \centering
  \begin{tabular}{ cccc } 
    \includegraphics[width=0.25\linewidth]{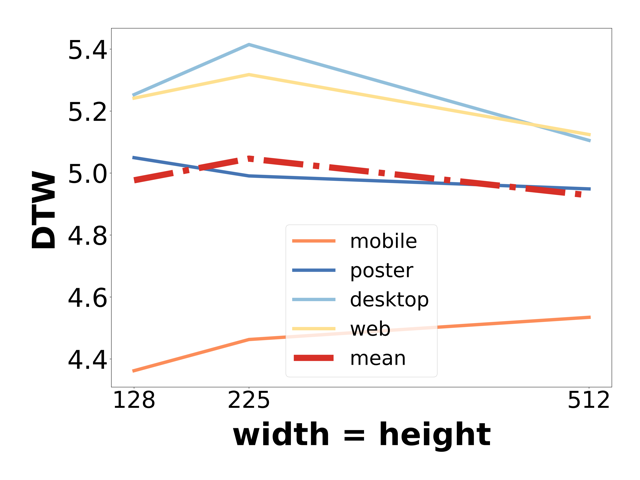}   \includegraphics[width=0.25\linewidth]{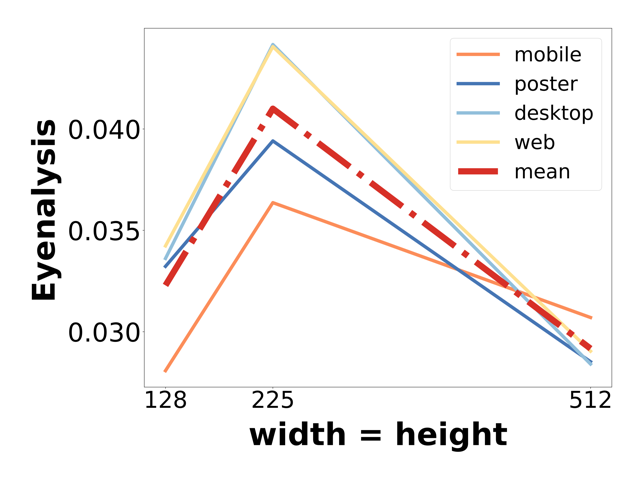}
    \includegraphics[width=0.25\linewidth]{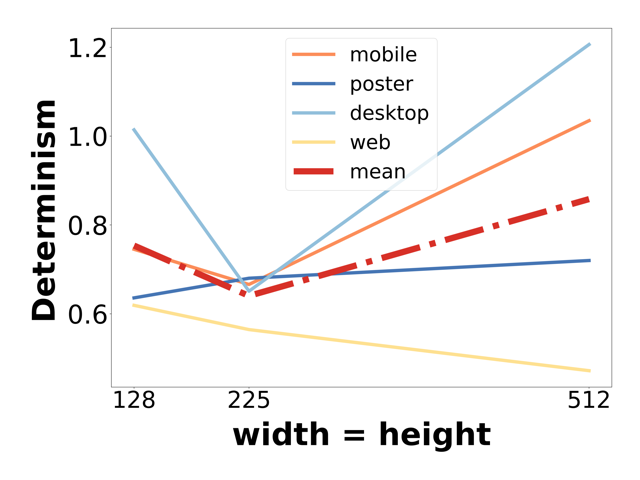} 
    \includegraphics[width=0.25\linewidth]{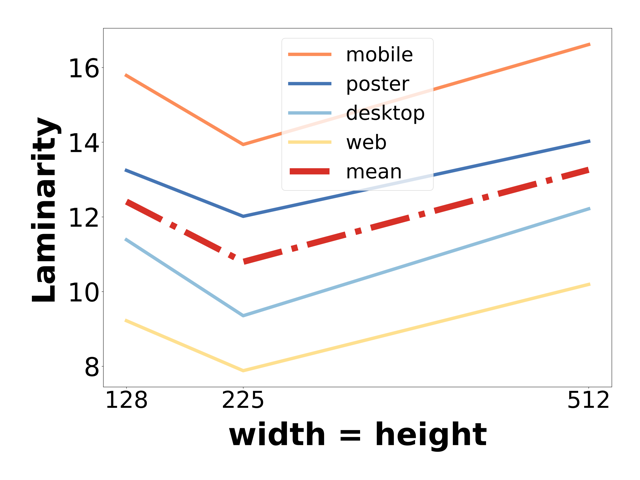}
    \\ 
  \end{tabular}
  \caption{
    Impact of resizing to different square image sizes on different GUI types. 
    We consider sizes of 128, 225, and 512\,px. 
    The best results are usually observed for the 128\,px cases.
  }
  \label{fig:sen-image-resizing}
\end{figure}

In the second experiment (\autoref{fig:sen-image-resizing}), we resized images down to various dimensions,
while ensuring a square aspect ratio, as per the results of our previous experiment.
The results in this case indicate that resizing images to smaller dimensions has a positive impact on the prediction of both scanpaths and fixation points in mobile UIs. However, the opposite holds true for desktop UIs, as they typically have smaller elements as compared to mobile UIs.

\subsection{Sensitivity to IOR decay}

\begin{figure}[!t]
  \centering
  \begin{tabular}{ cccc } 
    \includegraphics[width=0.25\linewidth]{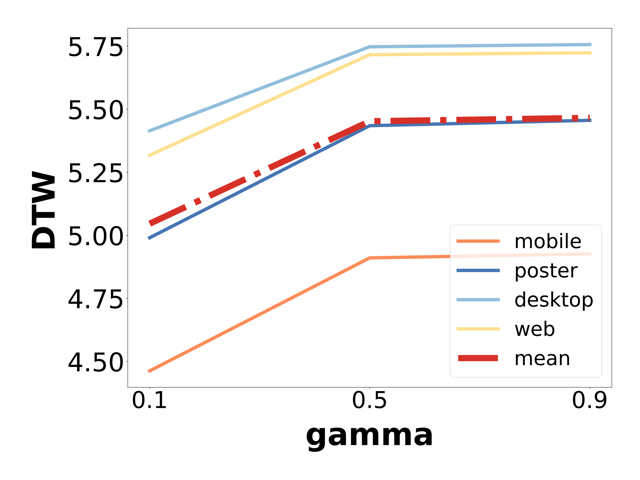}  \includegraphics[width=0.25\linewidth]{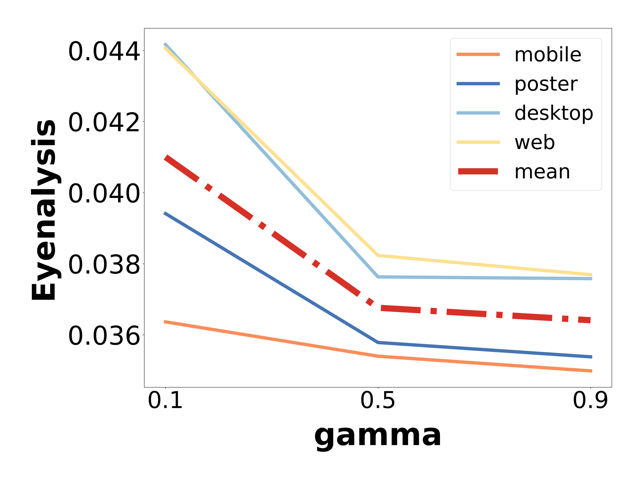}
    \includegraphics[width=0.25\linewidth]{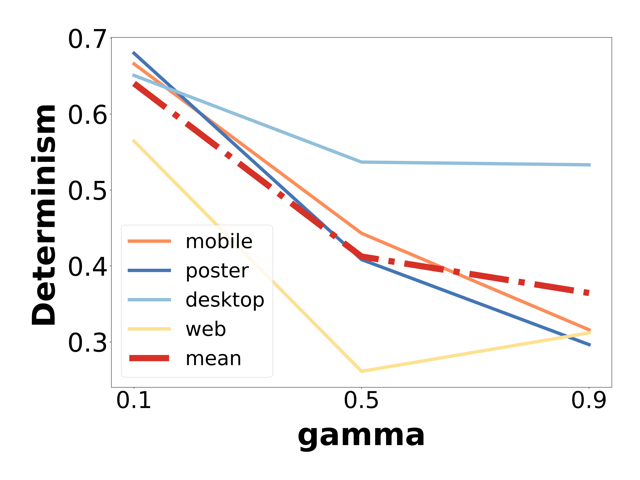} \includegraphics[width=0.25\linewidth]{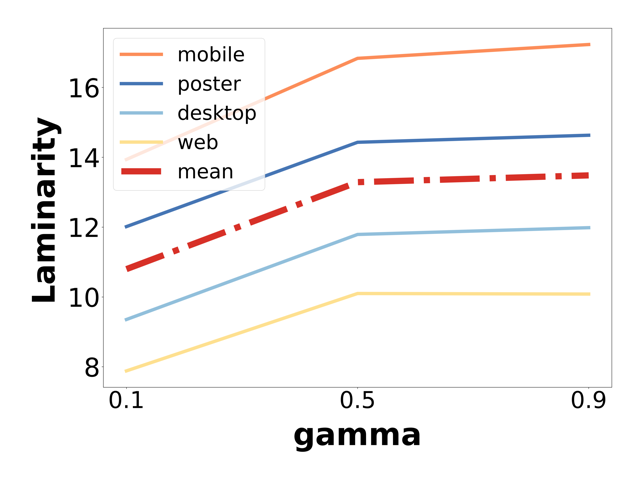} 
    \\ 
  \end{tabular}
  \caption{
    Impact of different $\gamma$ values on different GUI types. 
    Lower $\gamma$ means a high probability of revisiting fixation points.
    The best results are observed when $\gamma=0.1$.
    }
  \label{fig:sen-gamma}
\end{figure}

In this experiment, we varied the $\gamma$ parameter of our proposed IOR decay to assess its impact on scanpath prediction. 
As a reminder, a larger $\gamma$ indicates a higher probability of revisiting previously observed fixation points. 
\autoref{fig:sen-gamma} illustrates the findings of this experiment, 
indicating that smaller $\gamma$ values lead to improved scanpath prediction performance. 
This suggests that when the likelihood of revisiting a previously observed fixation point is low, 
the model performs better in predicting subsequent fixation points. 
Conversely, when the likelihood of revisiting a fixation point is high, the model excels in predicting individual fixation points.

\subsection{Sensitivity to masking radius}

\begin{figure}[!ht]
  \centering
  \begin{tabular}{ cccc } 
    \includegraphics[width=0.25\linewidth]{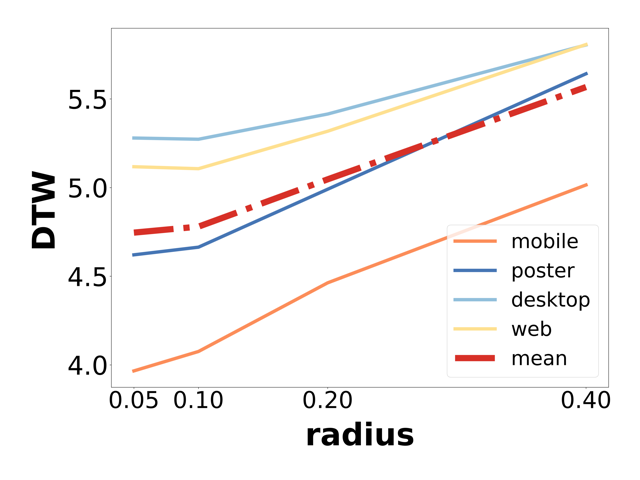}  \includegraphics[width=0.25\linewidth]{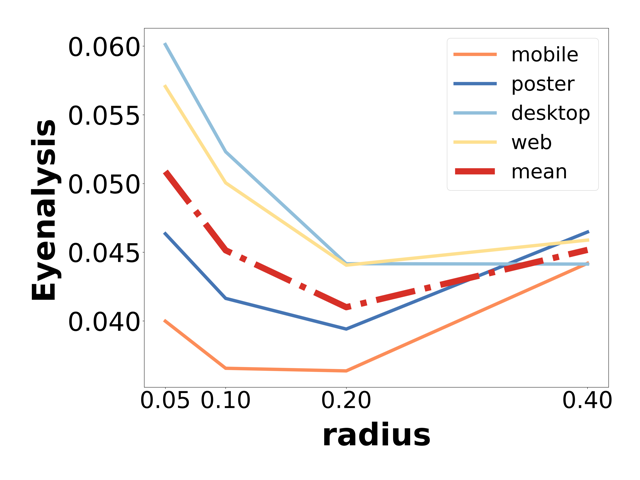} 
    \includegraphics[width=0.25\linewidth]{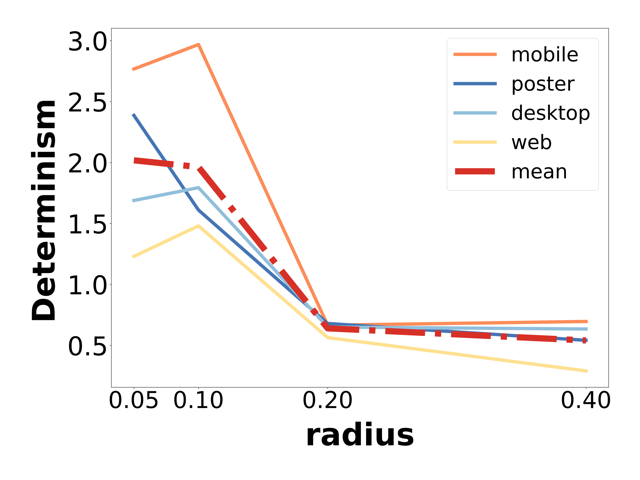} \includegraphics[width=0.25\linewidth]{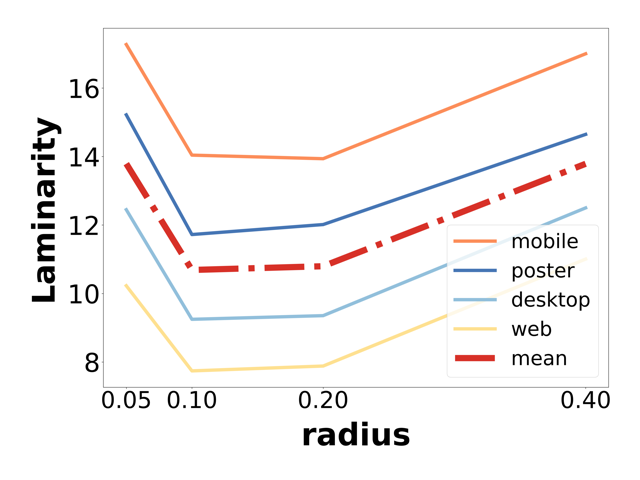} 
    \\ 
  \end{tabular}
  \caption{
    Impact of different masking radius on different GUI types.
    Masking radii are relative to the input image size (e.g. 0.2 means 20\% of the size).
    The best results are observed when the radius is set to 0.05.
    }
  \label{fig:sen-radius}
\end{figure}

In this experiment, we examined how altering the masking radius impacts scanpath prediction performance. 
The results are provided in \autoref{fig:sen-radius}.
We observed a negative correlation between the masking radius and the quality of the scanpath predictions, 
indicating that, as the radius increases, scanpath prediction quality decreases. 
However, we observed a sweet spot when the radius is set between 0.1 and 0.2, 
as better results are obtained according to the three non-DTW metrics.

\subsection{Putting it all together}

\begin{figure}[!ht]
  \centering
  \begin{tabular}{ cccc }
    \includegraphics[width=0.25\linewidth]{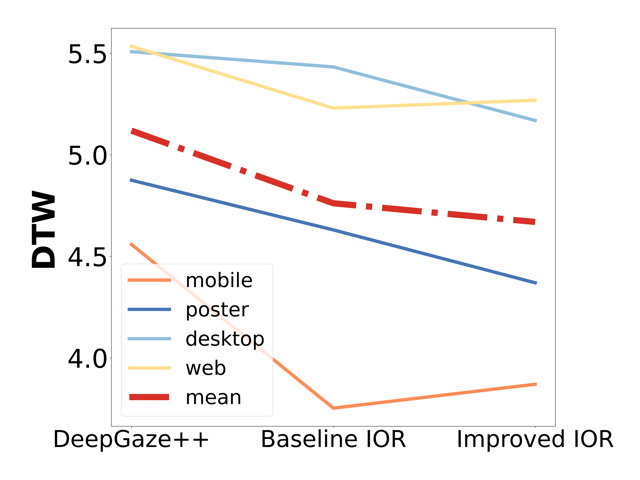}    \includegraphics[width=0.25\linewidth]{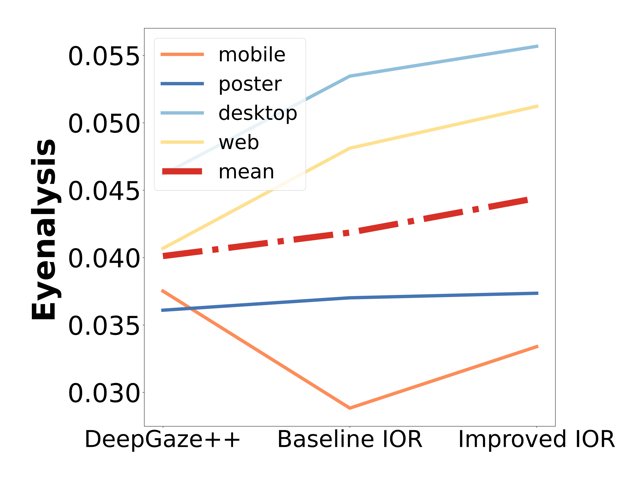}
    \includegraphics[width=0.25\linewidth]{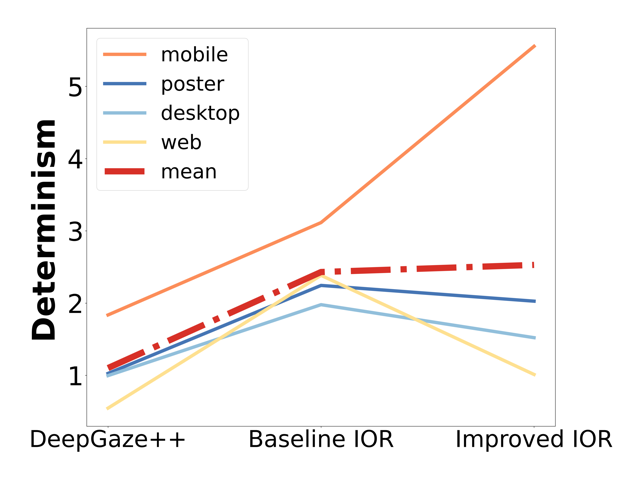}  \includegraphics[width=0.25\linewidth]{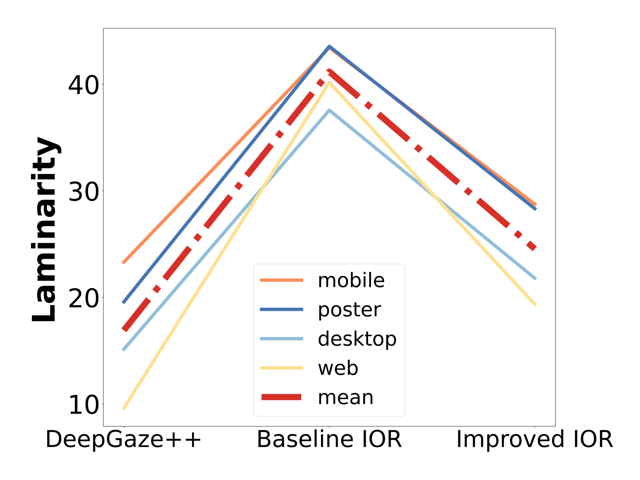} 
    \\ 
  \end{tabular}
  \caption{
    Impact of different IOR mechanisms, using optimal parameters, on different GUI types.
    ```Baseline IOR'' uses DeepGaze++ with the original IOR decay and the optimal parameters. 
    ```Improved IOR'' uses DeepGaze++ with our proposed IOR decay and the optimal parameters.
    The best results are usually observed with the baseline IOR. 
  }
  \label{fig:sen-ior}
\end{figure}

With the optimal parameters in place, we conducted an additional experiment on the test partition of UEyes
to understand the impact of an improved scanpath model.
\autoref{fig:sen-ior} illustrates the results.
The ``DeepGaze++'' cases represent the baseline model implementation~\cite{jiang2023ueyes}.
The ``Baseline IOR'' cases represent DeepGaze++ using the original IOR decay 
and the optimal parameters derived from our experiments,
whereas the ``Improved IOR'' cases represent DeepGaze++ with our proposed IOR decay 
and the optimal parameters derived from our experiments.
The results highlight that adopting the new IOR decay  
addresses the challenge of the limited number of fixation points 
and contributes to enhanced prediction performance as compared with the baseline DeepGaze++ model,
although the baseline IOR with optimal parameters is comparable in predicting fixation points.

The results indicate significant improvements when using these optimal parameters, 
underscoring their substantial impact on prediction performance. 
\autoref{fig:visit_revisit}
provides additional evidence by showing the ratios of visited-revisited elements for three types of GUI elements
(image, text, face) following previous work~\cite{Leiva20_saliency, jiang2023ueyes}.

\autoref{tab:res-comparison} shows that, by setting all the optimized parameters, 
the results of DeepGaze++ improve in all the metrics except Eyenalysis. 
The table presents the results of DeepGaze++ with baseline parameters, as described in \cite{jiang2023ueyes},
and with the set of optimized parameters.
According to the two-sample paired $t$-test,
differences are statistically significant for all metrics except Eyenalysis:
\textbf{DTW}: $t(107)=5.36, p<.0001, d=0.367$;
\textbf{Eyenalysis}: $t(107)=5.36, p=.4503\ (n.s.), d=0.074$;
\textbf{Determinism}: $t(107)=3.60, p<.001, d=0.432$;
\textbf{Laminarity}: $t(107)=8.98, p<.0001, d=0.580$.
Effect sizes (Cohen's $d$) suggest a moderate practical importance of the results~\cite{Lakens13}.

\begin{table}[!t]
    \centering
    \begin{tabular}{*5c}
    \toprule
    \textbf{DeepGaze++} & \textbf{DTW}$\downarrow$ & \textbf{Eyenalysis}$\downarrow$ & \textbf{Determinism}$\uparrow$ & \textbf{Laminarity}$\uparrow$ \\
\midrule
Baseline
& $5.118\pm0.482$ &
$\mathbf{0.040\pm0.004}$ &
$1.101\pm0.536$ & 
$16.908\pm5.900$ \\
Improved & $\mathbf{4.669\pm0.667}$ &
$0.044\pm0.010$ & $\mathbf{2.529\pm2.059}$ & $\mathbf{24.557\pm4.705}$ \\
\bottomrule
    \end{tabular}
    \caption{Evaluation of baseline (original) and improved DeepGaze++ model (using the optimized parameters), 
    showing Mean $\pm$ SD results for each metric. 
    Arrows denote the direction of the importance; e.g., $\downarrow$ means “lower is better.” 
    Each column’s best result is highlighted in boldface. 
    }
    \label{tab:res-comparison}
\end{table}

\subsection{Analysis of visited and revisited patterns}

\begin{figure}[!t]
  \centering
  \begin{tabular}{ ccc } 
    \includegraphics[width=0.2\linewidth]{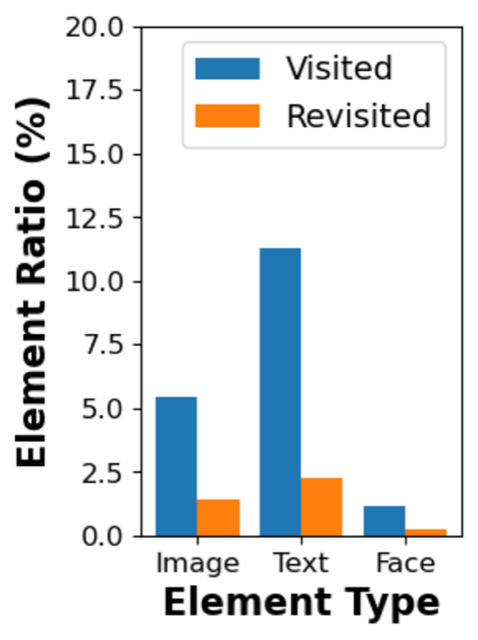}&  
    \includegraphics[width=0.2\linewidth]{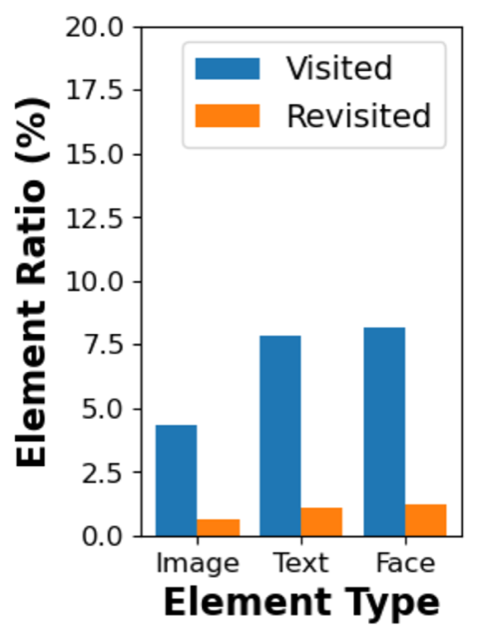}& 
    \includegraphics[width=0.2\linewidth]{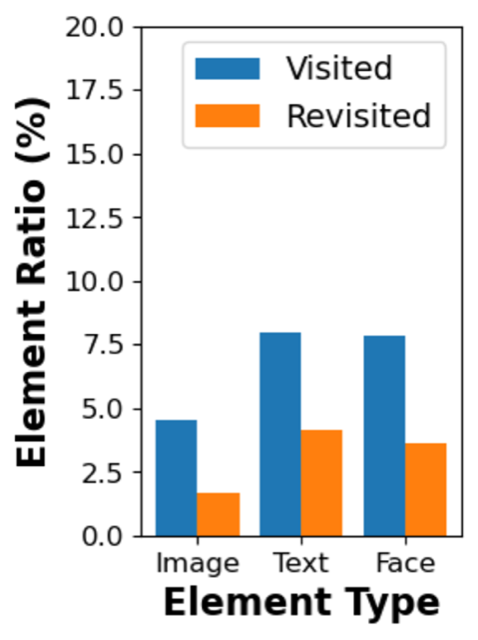}
    \\
    (a) Ground-truth &
    (b) Baseline DeepGaze++ &
    (c) Improved DeepGaze++
  \end{tabular}
  \caption{Visit vs. revisit bias analysis, showing the ratios of visited-revisited elements for three element categories. 
  According to the ground-truth data, text elements are more likely to be visited and revisited than images. 
  The improved model is better aligned with this observation.  
  }
  \label{fig:visit_revisit}
\end{figure}

In line with previous research that quantified the impact of scanpath models in GUI elements~\cite{jiang2023ueyes}, 
we categorized the GUI elements in UEyes into three categories (image, text, and face) 
using an enhanced version of the UIED model~\cite{xie2020uied}, which is designed to detect images and text on GUIs. 
We then analyzed the number of elements in each category that were initially visited and subsequently revisited. 
An element is considered revisited if the element gets a fixation again after at least three fixations on other elements.
The findings are presented in \autoref{fig:visit_revisit}.
We observed that text elements have a higher fixation probability than images in our improved model,
which is better aligned with the ground-truth cases. 
The improved model is also more aligned with the ground-truth cases in terms revisited fixations.
For visited fixations, no differences between models were observed.

\subsection{Example gallery}

\begin{figure}[!ht]
  \centering
  \small
  \begin{tabular}{ m{0.001\linewidth} m{0.2\linewidth}m{0.2\linewidth}m{0.2\linewidth}m{0.2\linewidth} m{0.001\linewidth} } 
  \toprule
    &\centering \textbf{Input image} & \centering \textbf{Ground-truth} & \centering \textbf{Baseline DeepGaze++} & \centering \textbf{Improved DeepGaze++} & 
    \\
    \toprule
     \begin{turn}{90} \centering    
\textbf{Mobile}
\end{turn} &
    \includegraphics[width=0.2\textwidth]{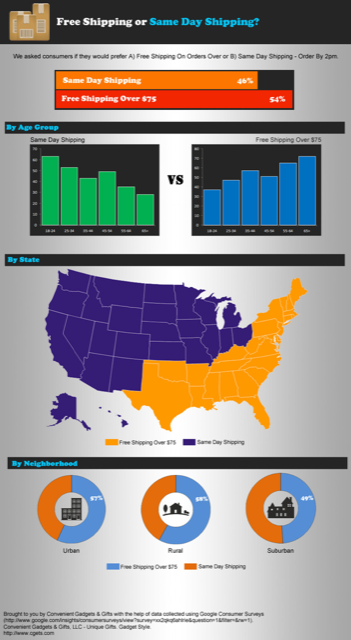} & 
    \includegraphics[width=0.2\textwidth]{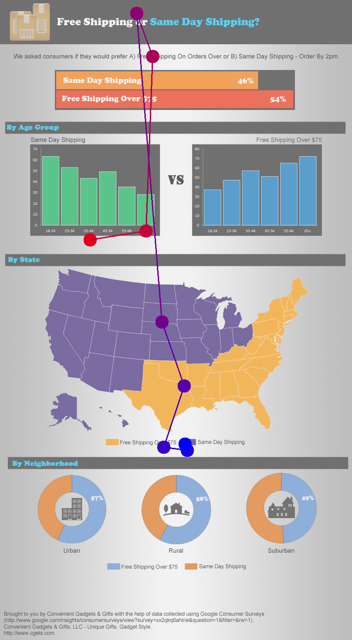} & 
    \includegraphics[width=0.2\textwidth]{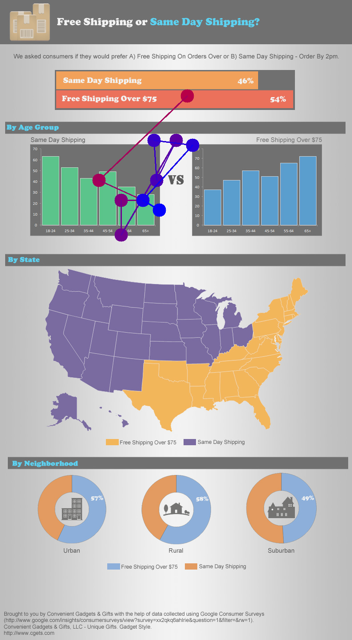} &
    \includegraphics[width=0.2\textwidth]{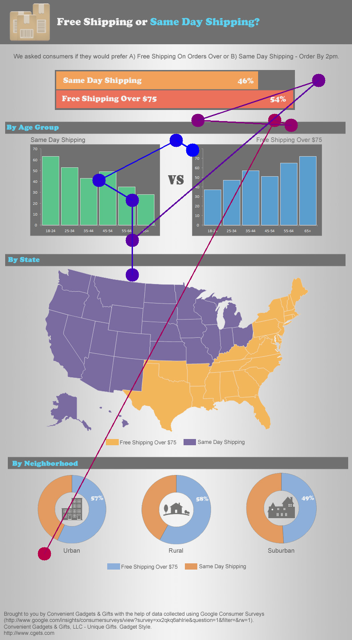}
    \\
    \toprule
     \begin{turn}{90} \centering
\textbf{Web} 
\end{turn} & 
    \includegraphics[width=0.2\textwidth]{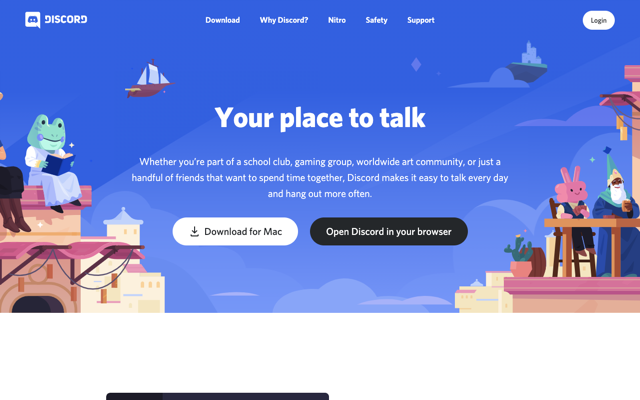} & 
    \includegraphics[width=0.2\textwidth]{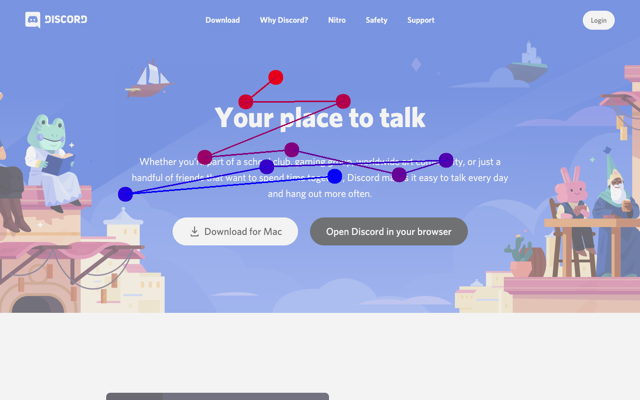} & 
    \includegraphics[width=0.2\textwidth]{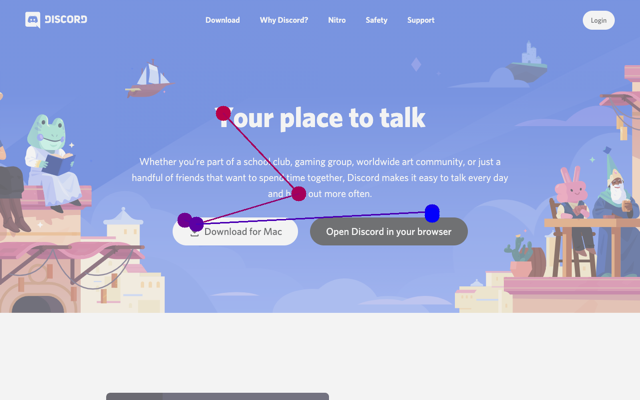} &
    \includegraphics[width=0.2\textwidth]{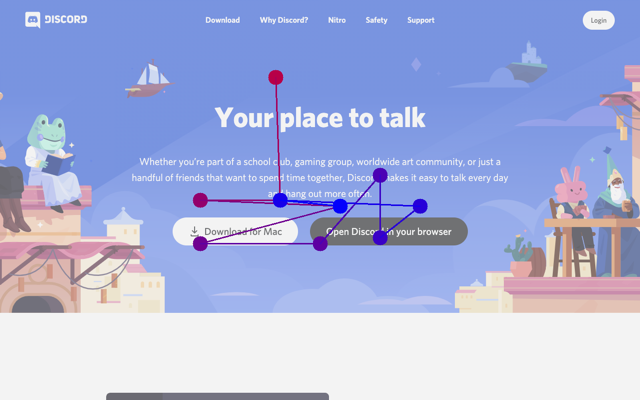}
    \\ 
    \toprule
    \begin{turn}{90} \centering
\textbf{Desktop} 
\end{turn} & 
    \includegraphics[width=0.2\textwidth]{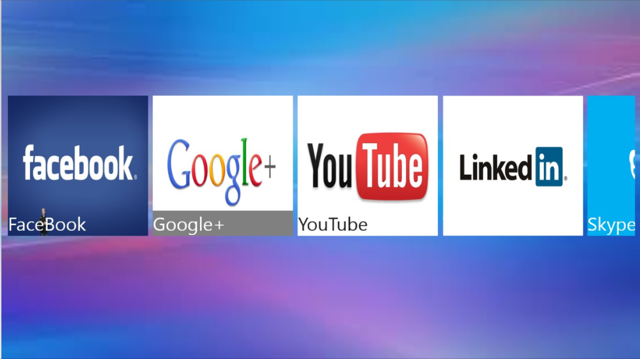} & 
    \includegraphics[width=0.2\textwidth]{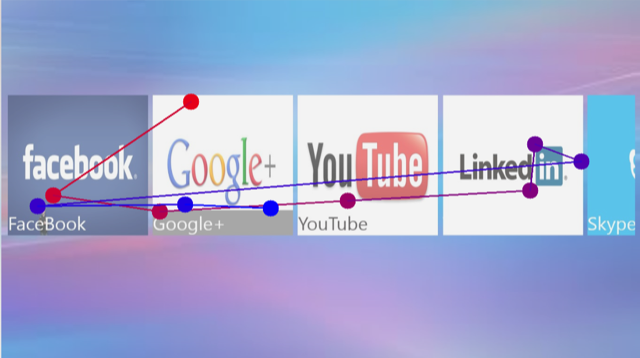} & 
    \includegraphics[width=0.2\textwidth]{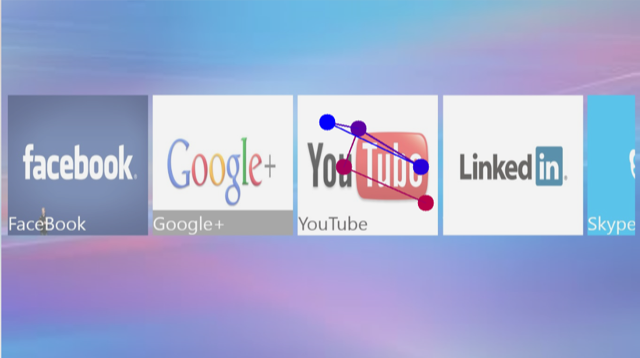} &
    \includegraphics[width=0.2\textwidth]{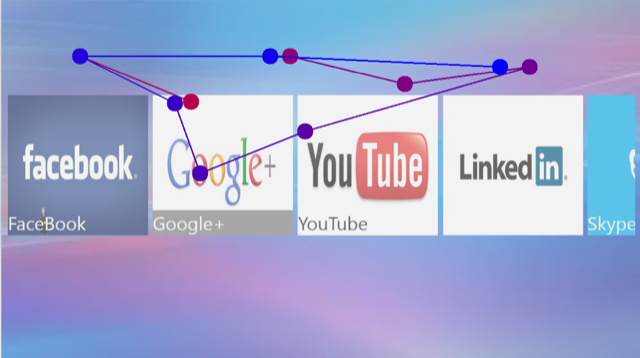}
    \\
    \toprule
    \begin{turn}{90} \centering
\textbf{Poster} 
\end{turn} & 
    \includegraphics[width=0.2\textwidth]{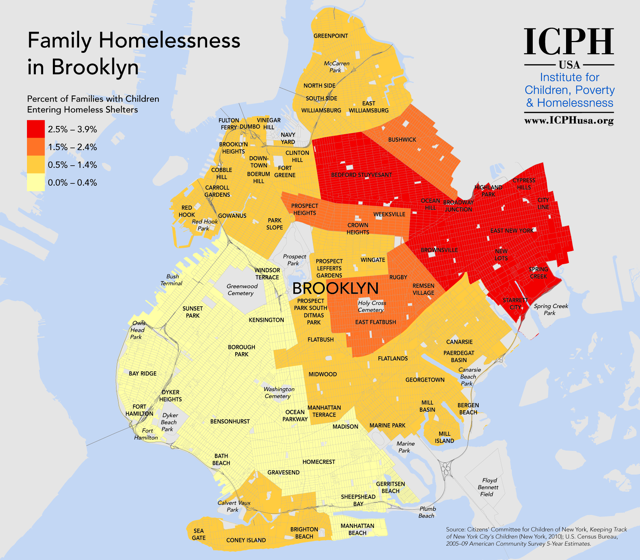} & 
    \includegraphics[width=0.2\textwidth]{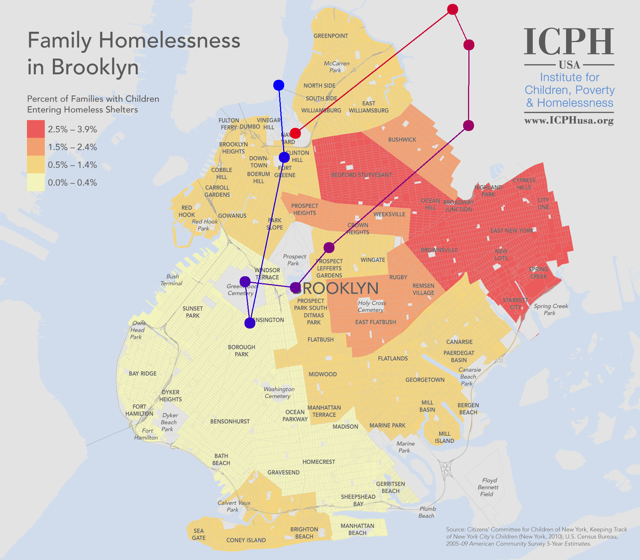} & 
    \includegraphics[width=0.2\textwidth]{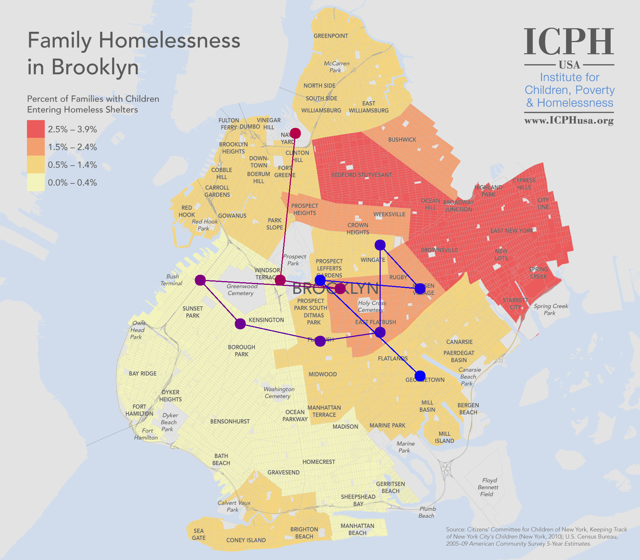} &
    \includegraphics[width=0.2\textwidth]{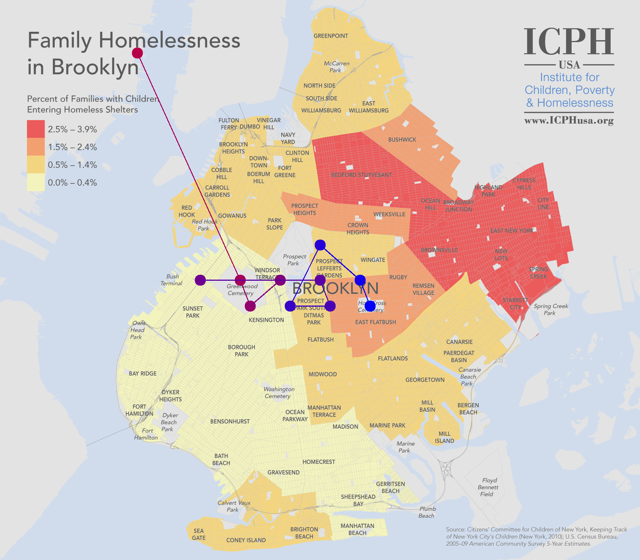} 
    \\
    \bottomrule
  \end{tabular}
  \caption{Qualitative comparison between scanpaths. 
  The scanpaths follow a color gradient from red (beginning of trajectory) to blue (end of trajectory).
  }
  \label{fig:gallery}
\end{figure}

\autoref{fig:gallery} illustrates our qualitative comparison of different scanpath models across various GUI types. 
The baseline DeepGaze++ model can predict fixation points but the resulting scanpaths are not very realistic. 
The improved DeepGaze++ model is able to predict realistic trajectories, with more accurate fixation points overall. 
It is worth noting that both models tend to have a center bias
and tend to generate clusters of fixation points.
The scanpaths shown in \autoref{fig:gallery} follow a color gradient
from red (beginning of trajectory) to blue (end of trajectory).

\begin{table}[!ht]
    \centering
    \begin{tabular}{ll *4c}
    \toprule
    \textbf{Model} & & \textbf{DTW}$\downarrow$ & \textbf{Eyenalysis}$\downarrow$ & \textbf{Determinism}$\uparrow$ & \textbf{Laminarity}$\uparrow$
    \\
\midrule

\multirow{2}{6em}{\textbf{Itti-Koch}} &
Baseline & $7.023\pm0.261$ & $0.075\pm0.014$ & $0.363\pm0.154$ & $\mathbf{5.823\pm1.169}$ 
\\
&
Improved & $\mathbf{5.824\pm0.219}$ & $\mathbf{0.053\pm0.012}$ & $\mathbf{0.378\pm0.141}$ & $4.943\pm1.034$
\\
\midrule
\multirow{2}{6em}{\textbf{Chen et al.}} &
Baseline & $4.298\pm0.225$  & $0.028\pm0.003$ & $\mathbf{1.597\pm1.556}$ & $7.028\pm0.344$
\\
&
Improved & $\mathbf{4.111\pm0.187}$ & $\mathbf{0.025\pm0.001}$ & $1.483\pm0.188$ & $\mathbf{7.724\pm1.259}$
\\
\midrule
\multirow{2}{6em}{\textbf{UMSS}} &
Baseline & $4.567\pm0.394$ & $\mathbf{0.031\pm0.006}$ &  $2.390\pm1.044$ & $10.541\pm1.764$
\\
&
Improved & $\mathbf{4.537\pm0.432}$ & $0.033\pm0.003$ & $\mathbf{3.123\pm1.028}$ & $\mathbf{12.302\pm1.815}$
\\
\midrule
\multirow{2}{6em}{\textbf{ScanGAN}} &
Baseline & $4.001\pm0.379$ & $\mathbf{0.026\pm0.003}$ & $1.306\pm1.025$ & $7.311\pm1.817$
\\
&
Improved & $\mathbf{3.973\pm0.204}$ & $0.027\pm0.002$ & $\mathbf{1.331\pm0.798}$ & $\mathbf{7.900\pm1.423}$
\\
\midrule
\multirow{2}{6em}{\textbf{ScanDMM}} &
Baseline & $4.584\pm0.336$ & $0.033\pm0.002$ & $\mathbf{0.597\pm0.499}$ & $5.123\pm1.042$
\\
&
Improved & $\mathbf{4.452\pm0.378}$ & $\mathbf{0.029\pm0.002}$ & $0.472\pm0.546$ & $\mathbf{5.596\pm0.996}$
\\
\bottomrule
\end{tabular}
    \caption{Evaluation of baseline and improved models, showing Mean $\pm$ SD for each metric. 
    Arrows denote the direction of the importance; e.g., $\downarrow$ means “lower is better.” 
    Each column’s best result is highlighted in boldface. 
    }
    \label{tab:comprehensive-comparison}
\end{table}

\begin{figure}[!ht]
  \centering
  \small
  \begin{tabular}{ m{0.001\linewidth} m{0.2\linewidth}m{0.2\linewidth}m{0.2\linewidth}m{0.2\linewidth} m{0.001\linewidth}} 
  \toprule
    & \centering \textbf{DTW}$\downarrow$ 
    & \centering \textbf{Eyenalysis}$\downarrow$ 
    & \centering \textbf{Determinism}$\uparrow$
    & \centering\textbf{Laminarity}$\uparrow$
    &
    \\
    \midrule
     \begin{turn}{90}
     \centering
\textbf{DeepGaze++}
\end{turn} &  
    \includegraphics[width=0.23\textwidth]{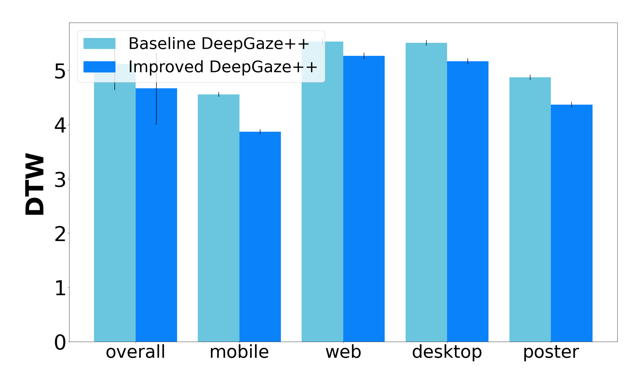} & 
    \includegraphics[width=0.23\textwidth]{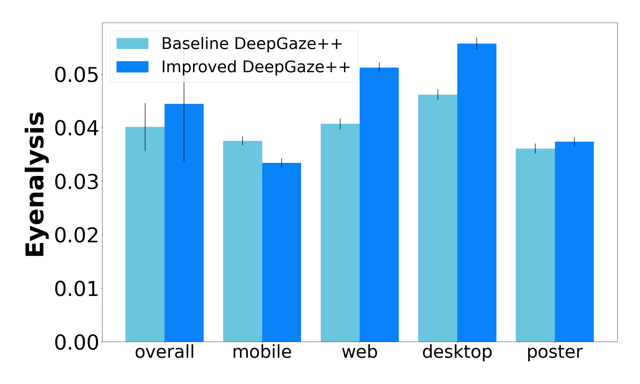} & 
    \includegraphics[width=0.23\textwidth]{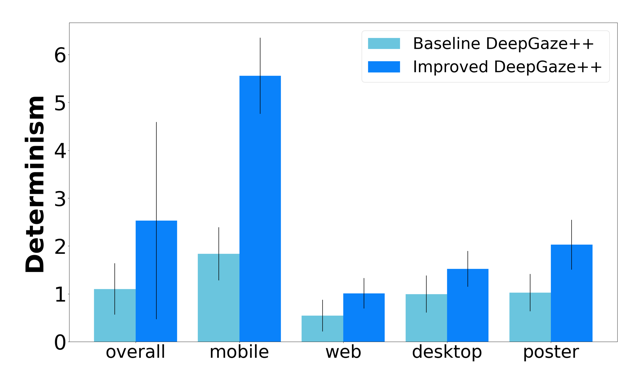} &
    \includegraphics[width=0.23\textwidth]{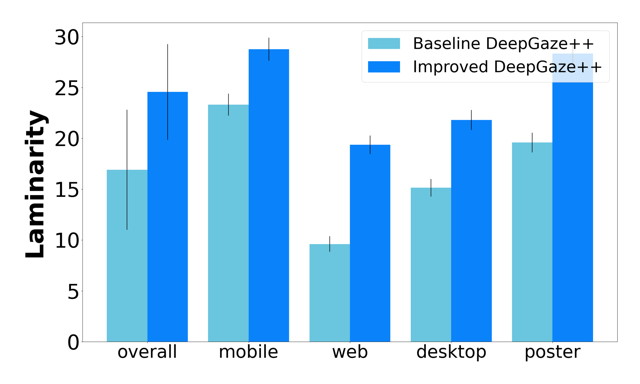}
    \\
    \midrule
     \begin{turn}{90} 
     \centering
\textbf{Itti-Koch} 
\end{turn} & 
    \includegraphics[width=0.23\textwidth]{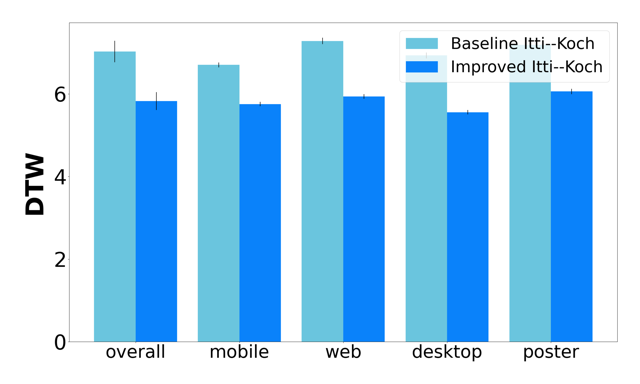} & 
    \includegraphics[width=0.23\textwidth]{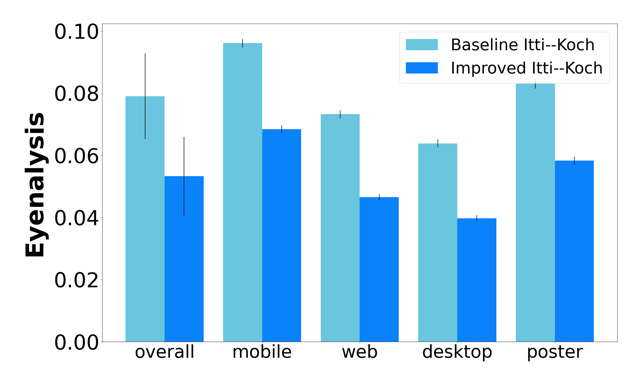} & 
    \includegraphics[width=0.23\textwidth]{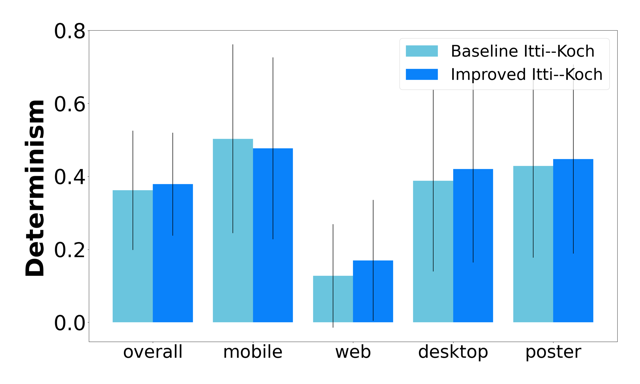} &
    \includegraphics[width=0.23\textwidth]{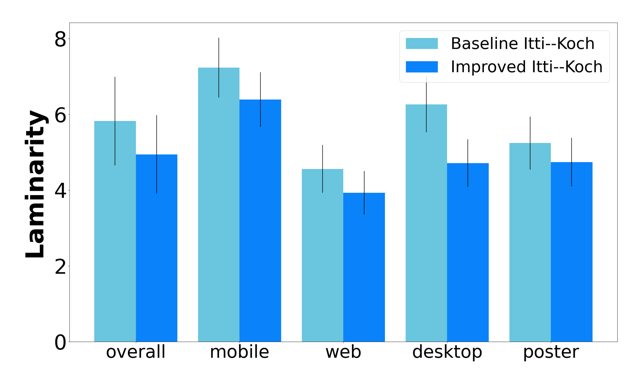}
    \\ 
    \midrule
    \begin{turn}{90} 
    \centering
\textbf{Chen et al.} 
\end{turn} & 
    \includegraphics[width=0.23\textwidth]{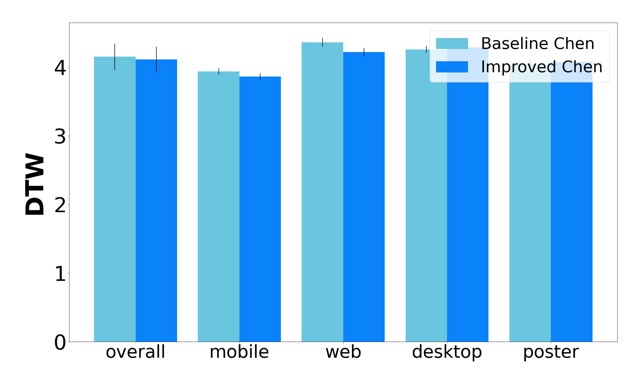} & 
    \includegraphics[width=0.23\textwidth]{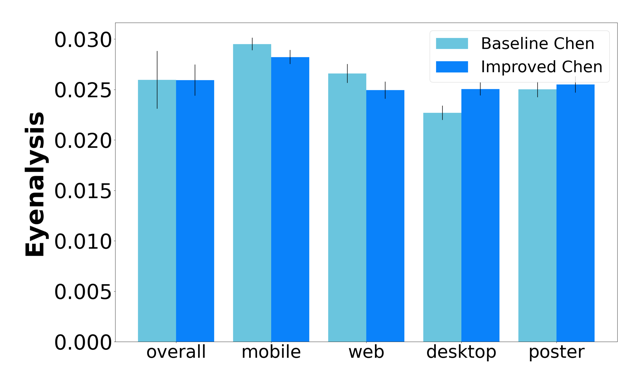} & 
    \includegraphics[width=0.23\textwidth]{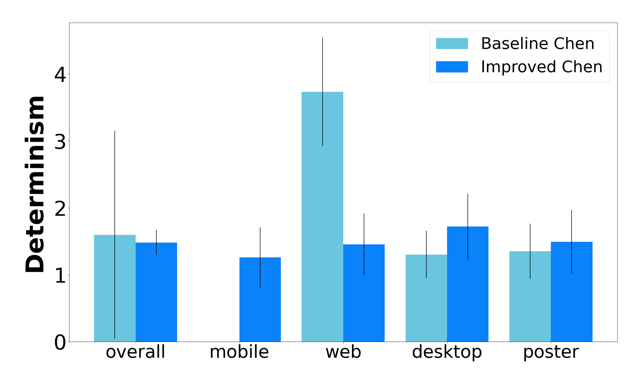} &
    \includegraphics[width=0.23\textwidth]{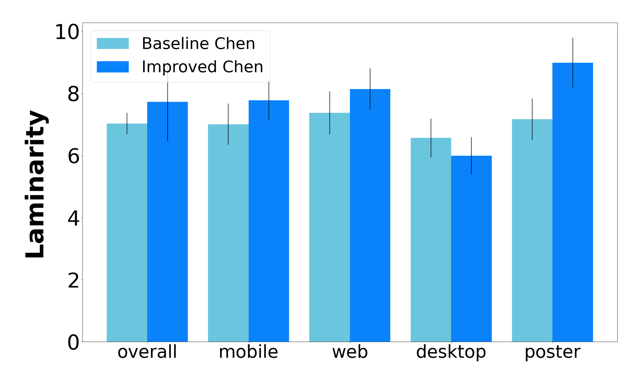}
    \\
    \midrule
    \begin{turn}{90} 
    \centering
\textbf{UMSS} 
\end{turn} & 
    \includegraphics[width=0.23\textwidth]{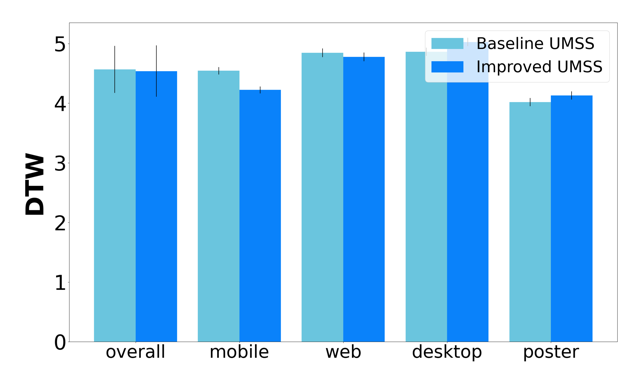} & 
    \includegraphics[width=0.23\textwidth]{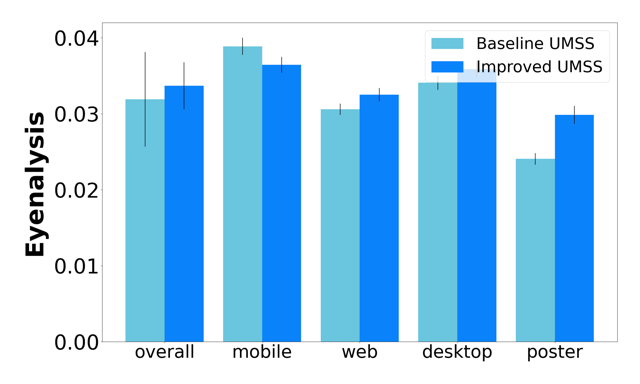} & 
    \includegraphics[width=0.23\textwidth]{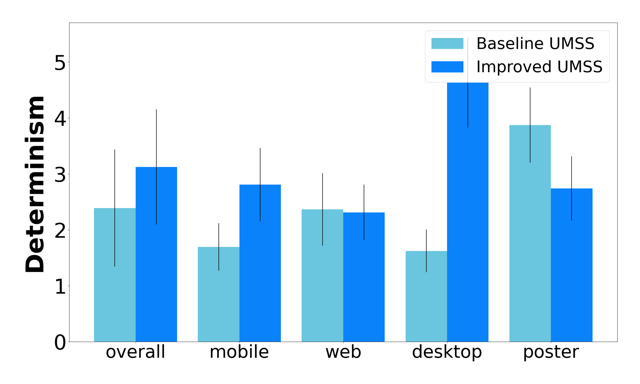} &
    \includegraphics[width=0.23\textwidth]{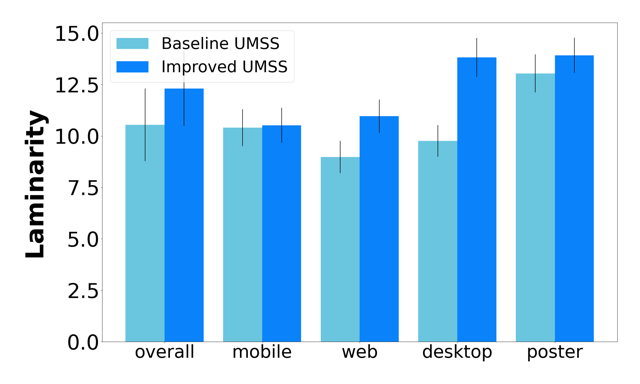}
     \\
    \midrule
    \begin{turn}{90} 
    \centering
\textbf{ScanGAN} 
\end{turn} & 
    \includegraphics[width=0.23\textwidth]{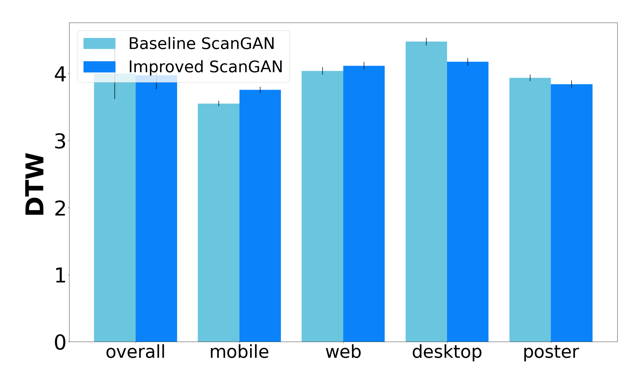} & 
    \includegraphics[width=0.23\textwidth]{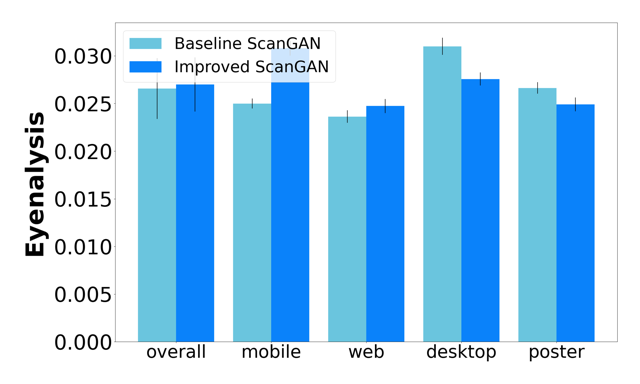} & 
    \includegraphics[width=0.23\textwidth]{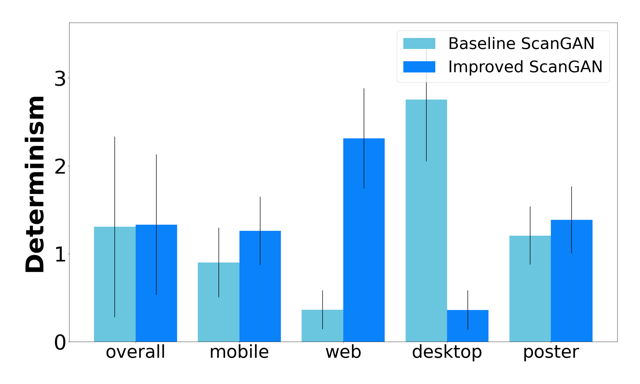} &
    \includegraphics[width=0.23\textwidth]{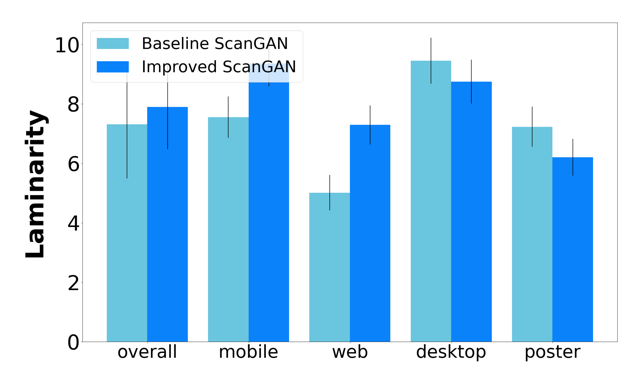}
     \\
    \midrule
    \begin{turn}{90} 
    \centering
\textbf{ScanDMM}
\end{turn} & 
    \includegraphics[width=0.23\textwidth]{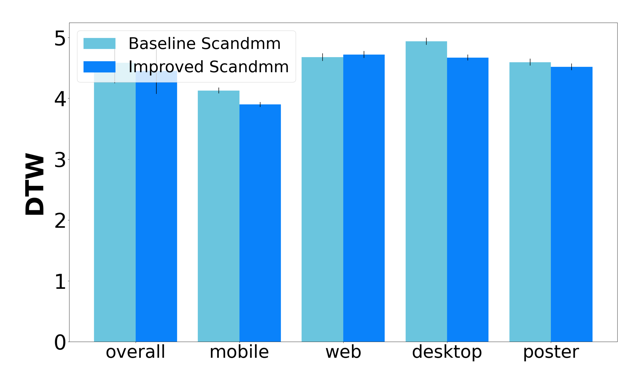} & 
    \includegraphics[width=0.23\textwidth]{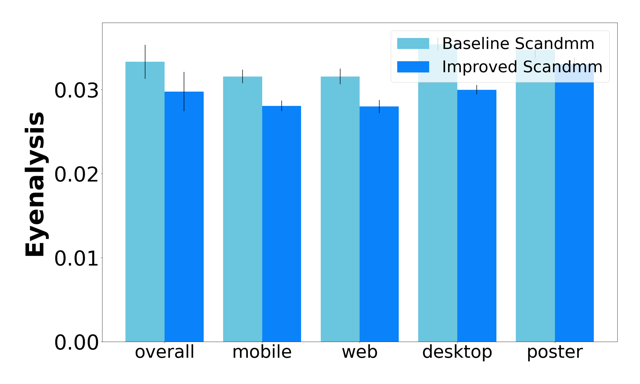} & 
    \includegraphics[width=0.23\textwidth]{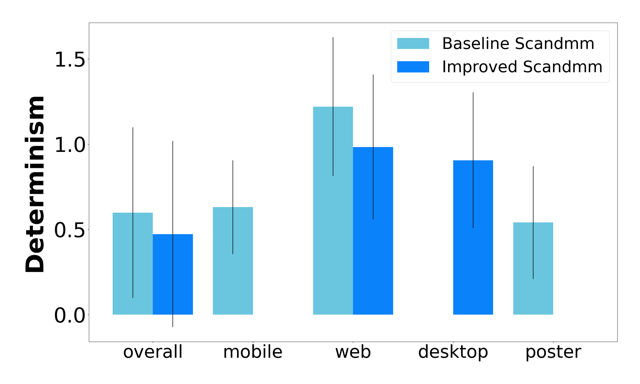} &
    \includegraphics[width=0.23\textwidth]{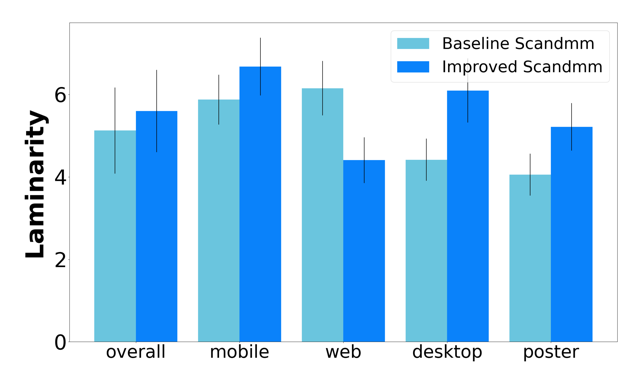}
    \\
    \bottomrule
  \end{tabular}
  \caption{Comparison against other scanpath models across GUI types.
  Each model (baseline, light blue bars) is re-trained with the optimized parameters (improved, dark blue bars) 
  and evaluated on the testing partition.
  Error bars denote standard deviations}.
  \label{fig:multiple_model}
\end{figure}

\subsection{Comparison against other scanpath models}

To show the generalizability of our findings, we further conducted evaluations on a more diverse set of scanpath prediction models:
Itti-Koch model~\cite{itti1998model},
UMSS~\cite{wang2023scanpath},
ScanGAN~\cite{martin2022scangan360}, 
ScanDMM~\cite{Sui_2023_CVPR},
and the model by~\citet{chen2021predicting}.
We applied the same set of optimized parameters obtained from DeepGaze++ to these other models. 
\autoref{fig:multiple_model} and \autoref{tab:comprehensive-comparison} show the results.
This approach allowed us to assess the performance of the parameters across different scanpath models, 
providing valuable insights into their applicability beyond a single model.

All the models show improved results on most metrics after employing the optimized set of parameters on most types of GUIs.
Similar to DeepGaze++, the Itti-Koch model also incorporates the IOR mechanism.
Therefore, adjusting the masking radius to its optimal value has a notable impact on prediction performance. 
According to our findings, it is clear that there is potential to enhance the performance of scanpath prediction models
by utilizing a set of optimal design parameters that cannot be learned from the data. 
This highlights the importance of considering and optimizing these design parameters 
to achieve improved performance in scanpath models.

\subsection{Comparison against other datasets}

To further demonstrate the generalization of our optimal parameters and the improved DeepGaze++ model, we evaluate it on MASSVIS~\cite{recall2016beyond}, one of the largest real-world visualization databases, scraped from various online sources including government reports, infographic blogs, news media websites, and scientific journals. MASSVIS includes scanpaths from 393 screenshots observed by 33 viewers, with an average of 16 viewers per visualization. Each viewer spent 10 seconds examining each visualization, resulting in an average of 37 fixation points. To accommodate the limitation of the baseline DeepGaze++ model of 12 fixation points, we considered the first 15 fixation points in each scanpath.

\autoref{tab:comprehensive-comparison_data} shows that the improved DeepGaze++ model consistently outperforms the baseline model on the four MASSVIS datasets across all scanpath metrics except Laminarity. 
When Laminarity is high but Determinism is low, it means that the scanpath model quantifies the number of locations that were fixated in detail in the ground-truth scanpath, but were only fixated briefly in the predicted scanpath~\cite{anderson2015comparison}.
In this regard, we can see that the improved model has always a smaller difference between these two metrics,
suggesting thus a better alignment with the ground-truth scanpaths.
Notably, the best improvements were observed on the InfoVis dataset and the best performance overall was observed on the Science dataset.

\begin{table}[!t]
    \centering
    \begin{tabular}{ll *4c}
    \toprule
    \textbf{Dataset} & & \textbf{DTW}$\downarrow$ & \textbf{Eyenalysis}$\downarrow$ & \textbf{Determinism}$\uparrow$ & \textbf{Laminarity}$\uparrow$
    \\
\midrule
\multirow{2}{6em}{\textbf{Government}} &
Baseline &  $8.073\pm1.932$ & $0.171\pm0.102$ & $0.253\pm4.057$ & $\mathbf{39.555\pm24.446}$
\\
&
Improved & $\mathbf{6.674\pm2.294}$ & $\mathbf{0.125\pm0.111}$ & $\mathbf{1.680\pm10.192}$ & $33.980\pm26.814$
\\
\midrule
\multirow{2}{6em}{\textbf{InfoVis}} &
Baseline & $7.318\pm1.782$ & $0.147\pm0.094$ & $1.418\pm10.334$ & $\mathbf{49.564\pm25.672}$
\\
&
Improved & $\mathbf{5.726\pm1.558}$ & $\mathbf{0.088\pm0.066}$ & $\mathbf{3.268\pm14.266}$ & $40.855\pm27.360$
\\
\midrule
\multirow{2}{6em}{\textbf{Science}} &
Baseline & $5.844\pm1.425$ & $0.074\pm0.040$ & $4.555\pm17.859$ & $\mathbf{49.658\pm26.855}$
\\
&
Improved & $\mathbf{5.323\pm1.475}$ & $\mathbf{0.064\pm0.047}$ & $\mathbf{5.611\pm18.521}$ & $45.203\pm25.027$
\\
\midrule
\multirow{2}{6em}{\textbf{News}} &
Baseline & $8.103\pm2.214$ & $\mathbf{0.163\pm0.134}$ & $0.110\pm2.926$ & $28.426\pm25.567$
\\
&
Improved & $\mathbf{7.648\pm2.776}$ & $0.168\pm0.169$ & $\mathbf{0.875\pm7.100}$ & $\mathbf{29.744\pm24.629}$
\\
\midrule
\multirow{2}{6em}{\textbf{Averaged}} &
Baseline & $7.334\pm1.058$ & $0.139\pm0.044$ & $1.584\pm2.065$ & $\mathbf{41.801\pm10.098}$
\\
&
Improved & $\mathbf{6.343\pm1.038}$ & $\mathbf{0.111\pm0.045}$ & $\mathbf{2.859\pm2.087}$ & $37.445\pm6.907$
\\
\bottomrule
\end{tabular}
    \caption{Evaluation of baseline and improved DeepGaze++ in the MASSVIS datasets, showing Mean $\pm$ SD for each metric. 
    Arrows denote the direction of the importance; e.g., $\downarrow$ means “lower is better.” 
    The best result in each case is highlighted in boldface.
    }
    \label{tab:comprehensive-comparison_data}
\end{table}

\subsection{Understanding the role of the number of fixation points}

\begin{figure}[!t]
  \centering
  \small
  \begin{tabular}{ m{0.001\linewidth} m{0.2\linewidth}m{0.2\linewidth}m{0.2\linewidth}m{0.2\linewidth} m{0.001\linewidth} } 
  \toprule
    &\centering \textbf{DTW}$\downarrow$ & \centering\textbf{Eyenalysis}$\downarrow$ & \centering\textbf{Determinism}$\uparrow$ & \centering\textbf{Laminarity}$\uparrow$ & 
    \\
    \midrule
     \begin{turn}{90} \centering 
\textbf{DeepGaze++}
\end{turn} &  
    \includegraphics[width=0.23\textwidth]{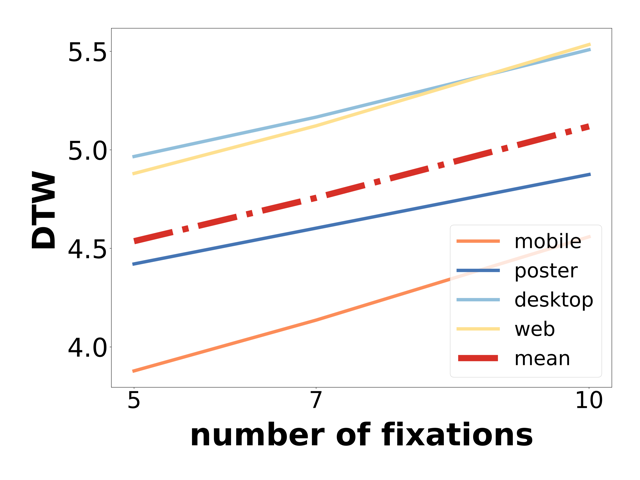} & 
    \includegraphics[width=0.23\textwidth]{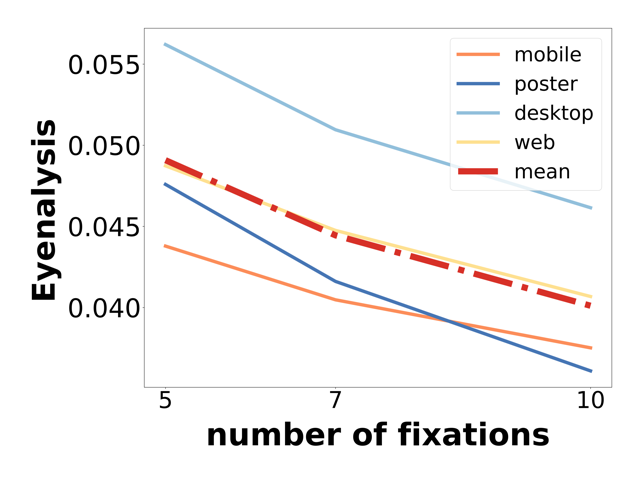} & 
    \includegraphics[width=0.23\textwidth]{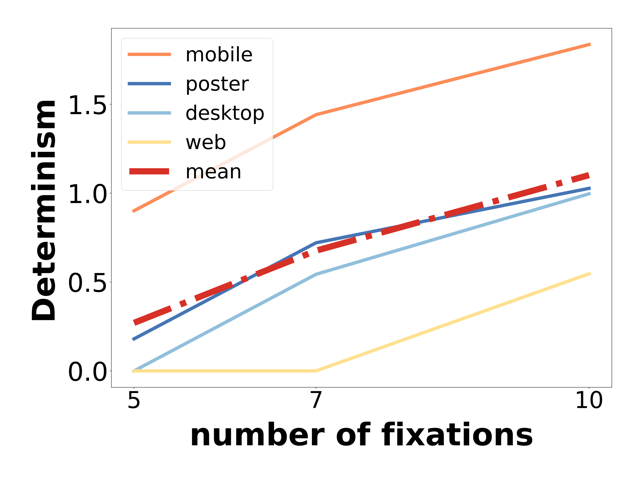} &
    \includegraphics[width=0.23\textwidth]{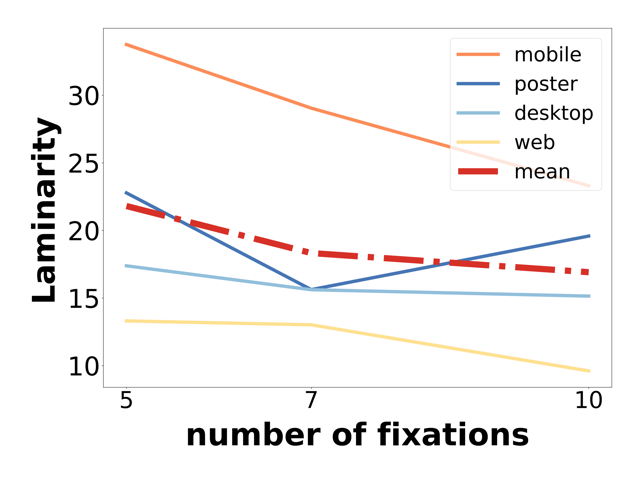}
    \\
    \midrule
     \begin{turn}{90} \centering
\textbf{Itti-Koch} 
\end{turn} & 
    \includegraphics[width=0.23\textwidth]{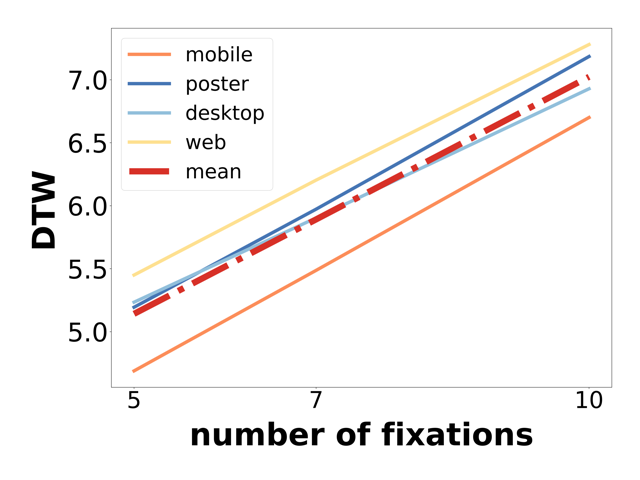} & 
    \includegraphics[width=0.23\textwidth]{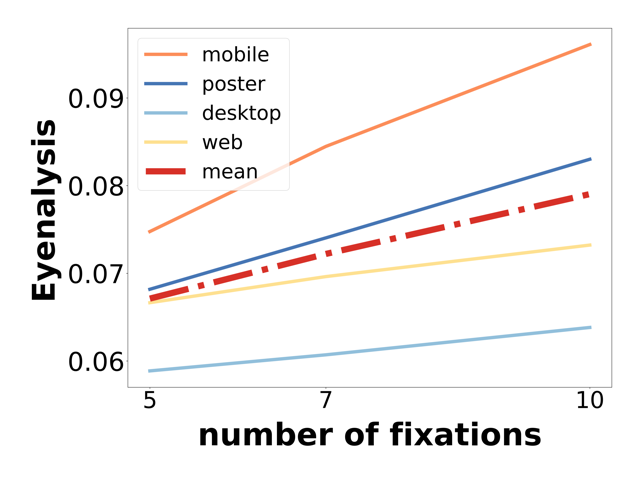} & 
    \includegraphics[width=0.23\textwidth]{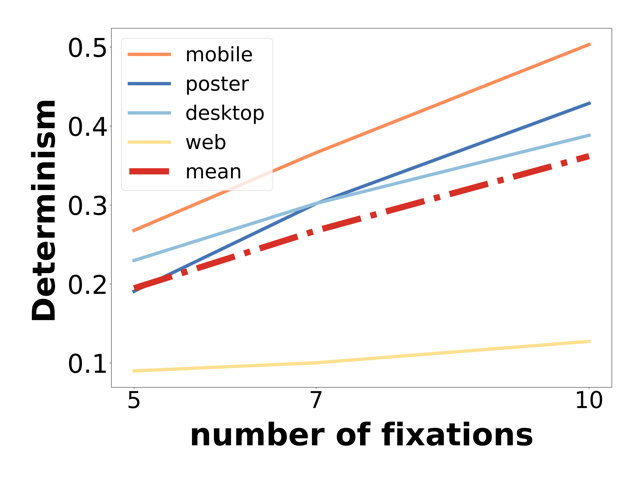} &
    \includegraphics[width=0.23\textwidth]{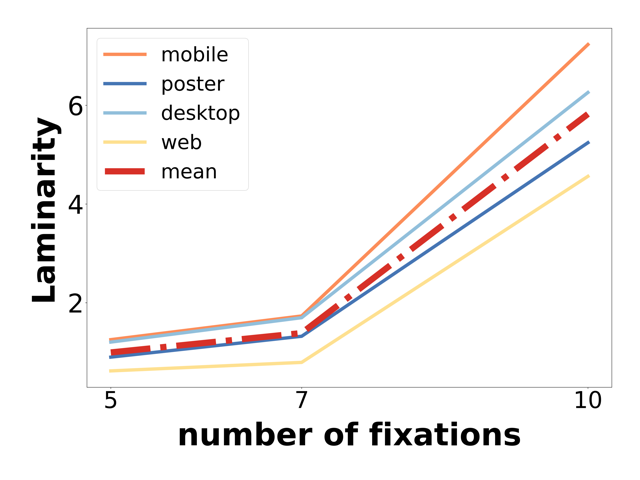}
    \\ 
    \midrule
    \begin{turn}{90} \centering
\textbf{Chen et al.} 
\end{turn} & 
    \includegraphics[width=0.23\textwidth]{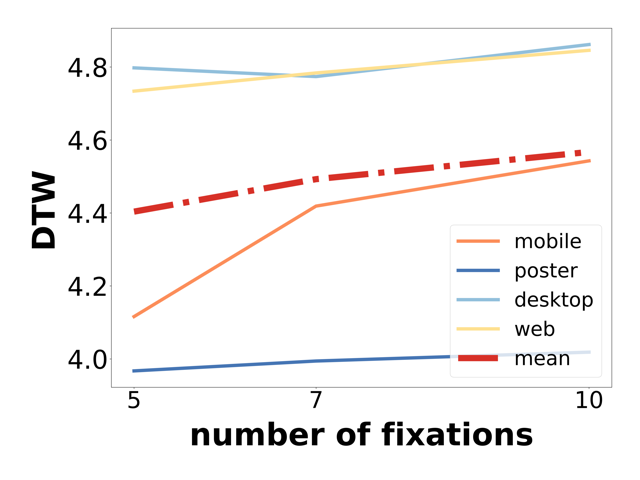} & 
    \includegraphics[width=0.23\textwidth]{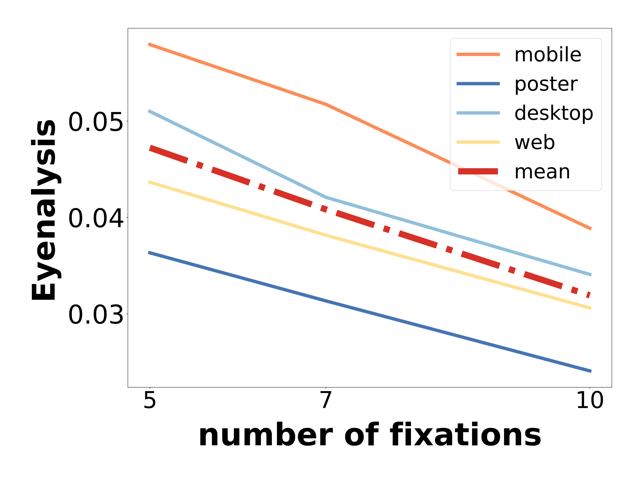} & 
    \includegraphics[width=0.23\textwidth]{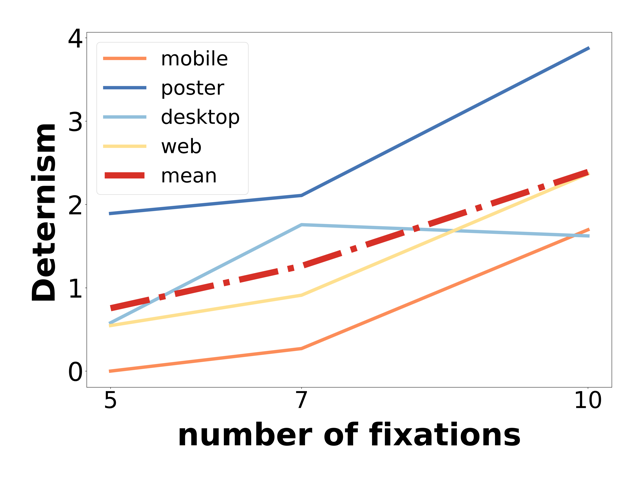} &
    \includegraphics[width=0.23\textwidth]{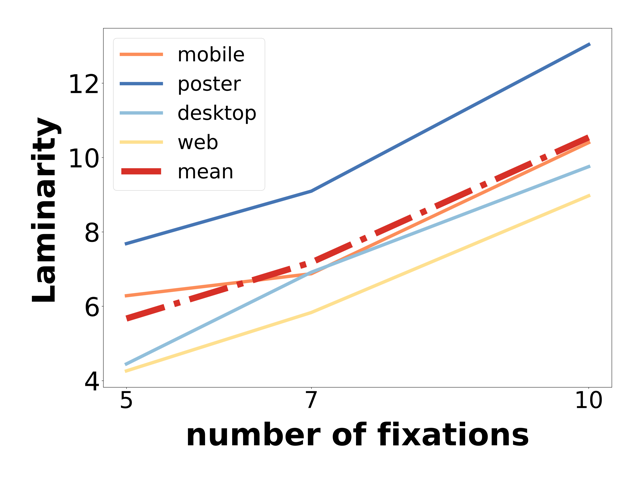}
    \\
    \midrule
    \begin{turn}{90} \centering
\textbf{UMSS} 
\end{turn} & 
    \includegraphics[width=0.23\textwidth]{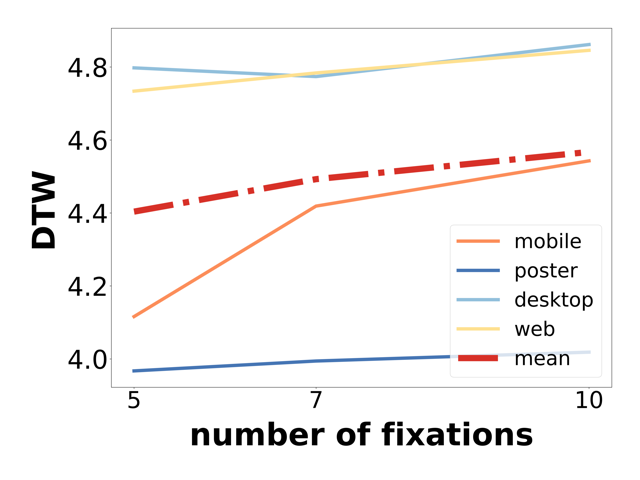} & 
    \includegraphics[width=0.23\textwidth]{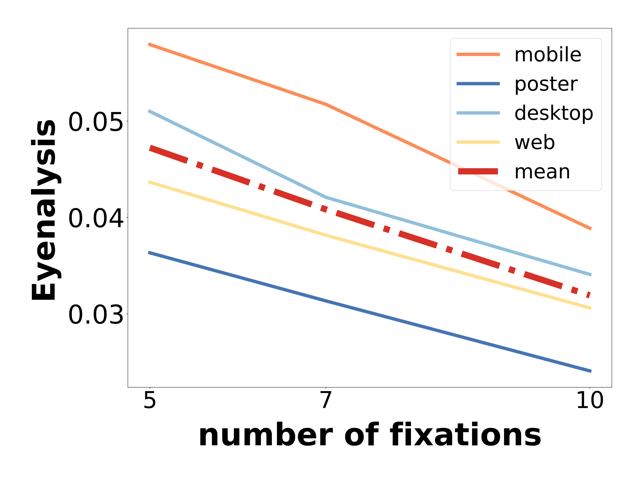} & 
    \includegraphics[width=0.23\textwidth]{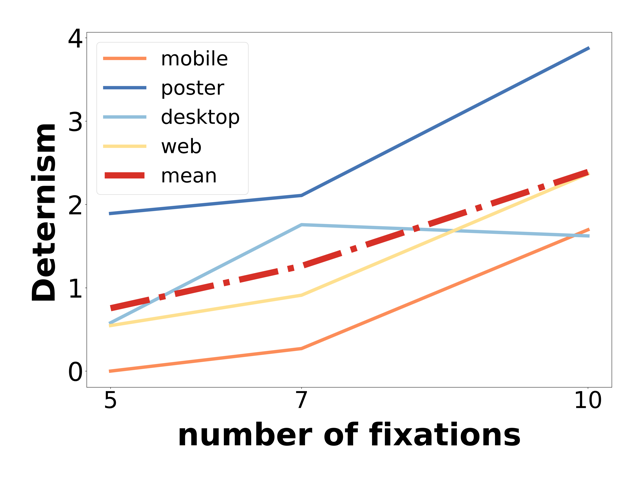} &
    \includegraphics[width=0.23\textwidth]{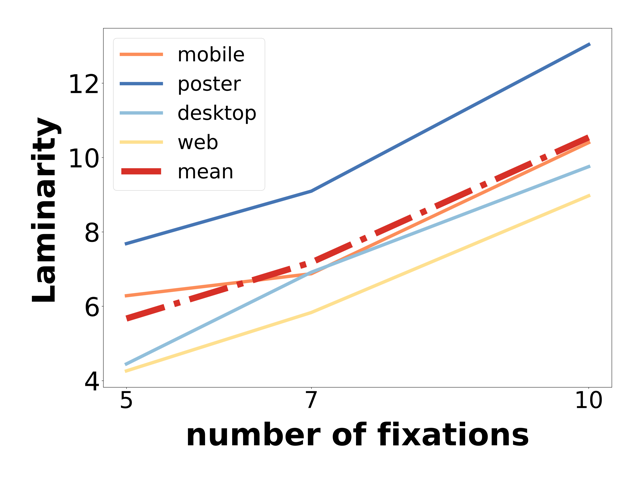}
     \\
    \midrule
    \begin{turn}{90} \centering
\textbf{ScanGAN} 
\end{turn} & 
    \includegraphics[width=0.23\textwidth]{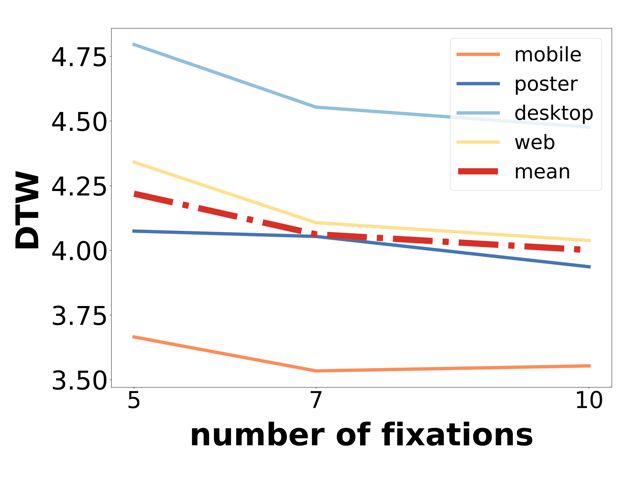} & 
    \includegraphics[width=0.23\textwidth]{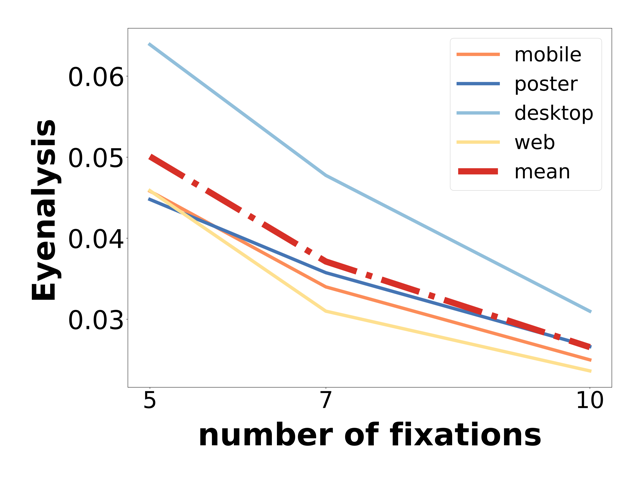} & 
    \includegraphics[width=0.23\textwidth]{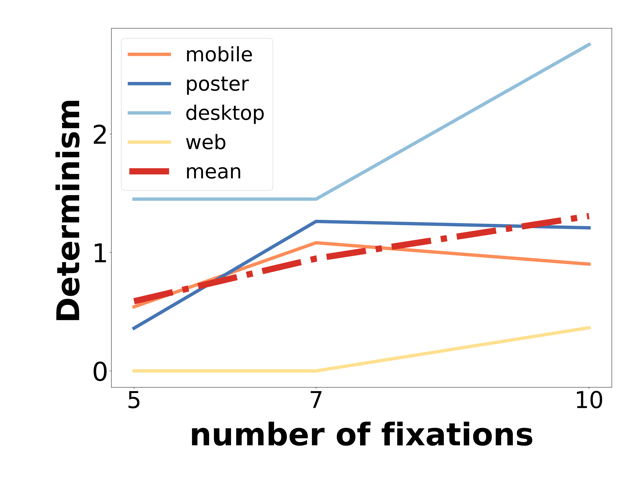} &
    \includegraphics[width=0.23\textwidth]{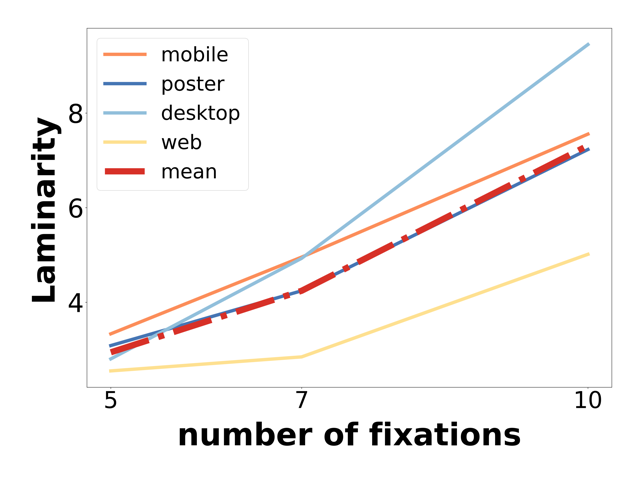}
     \\
    \midrule
    \begin{turn}{90} \centering 
\textbf{ScanDMM}
\end{turn} & 
    \includegraphics[width=0.23\textwidth]{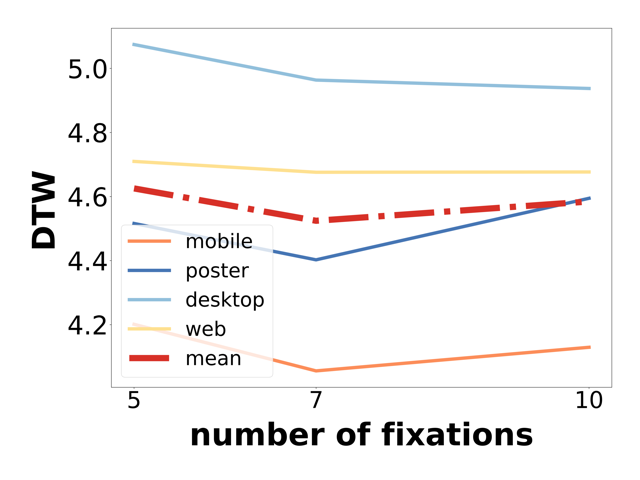} & 
    \includegraphics[width=0.23\textwidth]{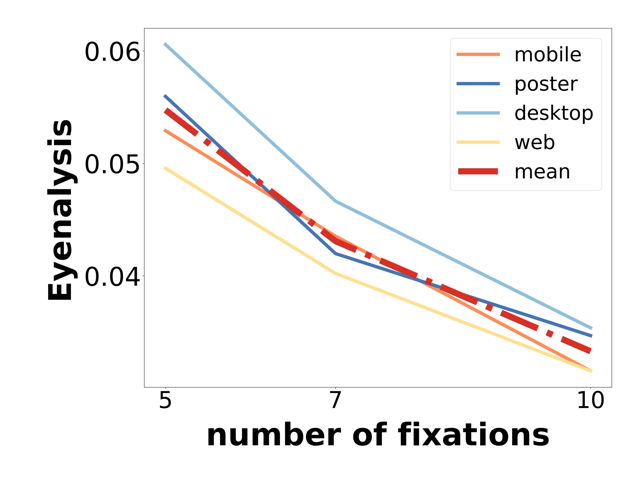} & 
    \includegraphics[width=0.23\textwidth]{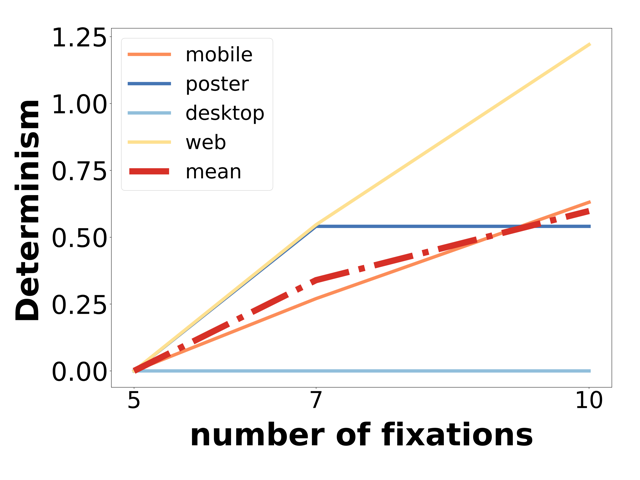} &
    \includegraphics[width=0.23\textwidth]{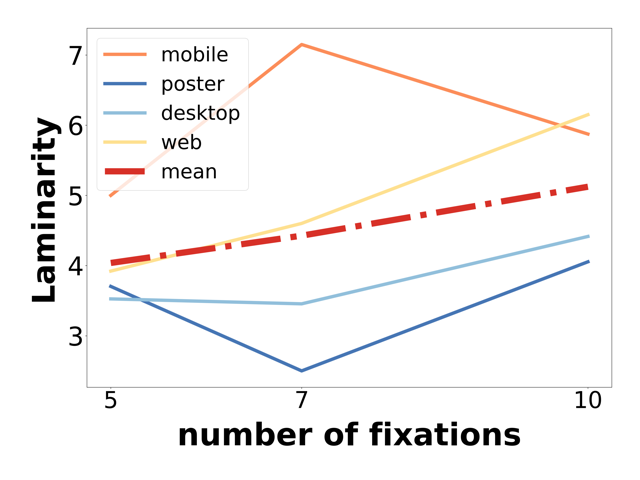}
    \\
    \bottomrule
  \end{tabular}
  \caption{Impact of different numbers of fixation points on different models on different GUI types.}
  \label{fig:number_of_fixation}
\end{figure}

All scanpath models are ultimately used to produce a number of fixation points.
While we do not consider this to be a design parameter, since it is actually a model outcome,
we do find it interesting to study their role in downstream performance.
Therefore, we conducted an additional analysis across all the scanpath models considered in our previous experiment. 
We systematically varied the number of fixation points from 5 to 10
and evaluated model performance on the testing partition.
The results are shown in \autoref{fig:number_of_fixation}.

We can observe that an increase in the number of fixation points correlates with improved Determinism and Laminarity values across all models. In addition, Eyenalysis exhibits enhancement in predictive accuracy for more fixation points except the Itti-Koch model. Thus, scanpaths with a larger number of fixation points are more likely to simulate the human's real scanpaths.

\section{Discussion}

Despite the development of scanpath prediction models for GUIs, 
the extent to which the design parameter choices influence saliency predictions performance has remained underexplored. 
We have conducted comprehensive experiments in this regard, 
using a state-of-the-art scanpath model as a reference.
By understanding the significance of these parameters, 
we contribute to the body of knowledge of how people look at GUIs 
and how to better develop models to predict it.

\emph{To what extent do saliency predictions depend on the choices made in design parameters?}
Our findings draw attention to the considerable influence of design parameters in determining the accuracy of predicting scanpaths in GUIs.
Specifically, the role of input image size, masking radius, and IOR decay is significant 
in assessing user attention and eye movement patterns in GUIs.
As shown in \autoref{fig:multiple_model}, optimizing these parameters can substantially enhance scanpath prediction performance. 
In summary, our research has led to the following findings:

\begin{enumerate}

\item Image size has a large impact on model predictions.
Resizing images to smaller dimensions positively impacts prediction performance
The best results were observed for images resized to 225\,px.

\item Resizing any input image to a square aspect ratio consistently yields superior performance across all GUI types. 
Mobile GUIs are particularly sensible to the image aspect ratio.

\item IOR is essential to reduce the likelihood of a user revisiting earlier seen GUI points. 
Our proposed decay $\gamma=0.1$ addresses an important limitation in DeepGaze++ and leads to improved prediction performance.

\item When the masking radius increases, prediction quality decreases.
The masking radius should find a balance between repetition of viewed areas (small radius) 
and blocking out too large parts of the GUI (large radius). 
A sensible value is $0.1$, i.e. 10\% of the available image size.

\item All the studied GUI types follow a similar trend in terms of optimal parameter settings,
although some GUIs may be affected slightly differently, as reported by the four evaluation metrics considered.

\end{enumerate}

Understanding the effect of design decisions on scanpath predictive models 
allows researchers to be aware of the fact that even small variations can lead to more accurate results. 
By examining how different design elements gauge users' attention,
researchers can identify effective design strategies that promote user engagement and optimize information presentation. 
This knowledge can be applied to various domains, including website design, multimedia content creation, and advertising, 
enabling designers to create more visually appealing and user-friendly interfaces. 
Furthermore, the evaluation on multiple scanpath models shows the generalizability of our findings.
We hope the insights presented in this paper could serve as a reference for future researchers working on saliency prediction in GUIs.

\subsection{Limitations and future work}

While our findings offer valuable new knowledge to optimize scanpath model performance, 
our experiments examined the impact of design parameters in isolation 
(i.e. we studied one design parameter at a time),
therefore future work should consider a joint optimization procedure.
It may be the case that an automatically optimized set (e.g. with Bayesian optimization)
can lead to more accurate performance results.

In principle, the proposed values of the design parameters we studied are meant to be applicable to every scanpath model. 
We found that this is the case for the 5 models we evaluated, most of them offering state-of-the-art performance, 
but we also acknowledge that there might be other sets of values that could work better for a particular model.

Future work should also consider more fixations points as model output.
In all our experiments, DeepGaze++ was used to predict trajectories of 10 fixations each, 
to facilitate comparisons against previous work~\cite{jiang2023ueyes, kummerer2022deepgaze}.
However, it may be the case that predicting more fixation points would result in more (or less) informed trajectories, 
which may in turn affect the performance evaluation metrics.
For example, if far-distant points (in time) tend to be more dispersed, the DTW values will increase. 
Such exploration can help develop better computational scanpath models.

\subsection{Privacy and ethics statement}

On the positive side, our research focuses on providing optimal parameters for scanpath models, enabling more accurate predictions. But this enhanced prediction goes beyond mere gaze direction, offering valuable insights into an individual's perceptual and cognitive processes. Our advancements open up opportunities for innovative applications, particularly in the realm of designing or adapting user interfaces. However, it is important to consider that the use of these optimal parameters and more accurate models can also be exploited for unforeseen purposes, such as optimizing advertisements placement on websites or enabling ``dark patterns'' such as making the user click on some content as a result of some GUI adaptation that optimized the interface elements for quick scannability. Overall, we should note that
striking a balance between harnessing the technology’s potential benefits and safeguarding individuals’ rights 
is crucial for responsible development and deployment.
\section{Conclusion} 

Scanpath models rely on a series of design parameters that are usually taken for granted.
We have shown that even small variations of these parameters have a noticeable impact on several evaluation metrics. 
As a result, we have found a set of optimal parameters that improve the state of the art in scanpath modeling.
These parameters have resulted in an improved DeepGaze++ model 
that can better capture both the spatial and temporal characteristics of scanpaths. 
These parameters are replicable to other computational models and datasets, 
showing the generalizability of our findings.
The community can use therefore this improved set of model parameters 
(or even the improved models themselves) 
to get a better understanding of how users are likely to perceive GUIs.
Ultimately, this work provides invaluable insights for designers and researchers interested in predicting users' viewing strategies on GUIs.
Our software and models are publicly available \url{https://github.com/prviin/scanpath-design-decisions}.

\begin{acks}
Research supported by the Horizon 2020 FET program of the European Union (grant CHIST-ERA-20-BCI-001)
and the European Innovation Council Pathfinder program (SYMBIOTIK project, grant 101071147).
\end{acks}

\bibliographystyle{ACM-Reference-Format}
\bibliography{Reference}

\end{document}